\magnification1200

\vskip 2cm
\centerline {\bf Local symmetry and extended spacetime}
\vskip 0.5cm

\centerline{Peter West}
\vskip 0.5cm
\centerline{{\it Mathematical Institute, University of Oxford,}}
\centerline{{\it Woodstock Road, Oxford, OX2 6GG, UK}}
\centerline{}
\centerline{{\it Department of Mathematics, King's College, London}}
\centerline{{\it The Strand, London WC2R 2LS, UK}}
\vskip 1cm
\centerline{ peter.west540@gmail.com}
\vskip 2cm
\leftline{\sl Abstract}  
We show that the previously proposed local transformations in E theory have a closing algebra if one adopts a simple constraint on the parameters of the transformations that constrains their dependence on the extended spacetime of E theory. We evaluate these local transformations for the decomposition  of $E_{11}$ that leads to the IIA theory  at level zero. This  has the same fields and spacetime as  Siegel theory (Double field theory) and although the  local transformations agree,  the  constraint is different to  the generic section conditions used  in that theory. We discuss what additional alternative constraints might be required for the invariance of this theory and also argue that the dependence on the coordinates beyond those of the usual spacetime is due to the presence of branes. 

\par 
\vskip2cm
\noindent

\vskip .5cm

\vfill
\eject

\medskip
{\bf 1 Introduction}
\medskip

Hidden symmetries in supergravity theories was first noticed in the context of the four dimensional $N=4$ supergravity theory which possess a   SL(2,R)/ U(1) symmetry [1]. The exceptional symmetries  [2,3,4]   appeared in the dimensional reduction of the supergravity theory in eleven dimensions [5]. The further the dimensional reduction was carried out the larger the symmetry  and  in particular the theory in four dimensions  possessed  an $E_7$ symmetry [2].  Although the eleven dimensional supergravity theory had no such symmetry, the IIA supergravity theory was found to have a  SO(1,1)  symmetry [6,7,8] and the IIB  supergravity theory [9,10,11] a SL(2,R)/ U(1) [9]. These symmetries were associated with the scalar fields of the supergravity theories  which belonged to a non-linear realisation of the symmetry.
\par
This construction of non linear realisations appeared long before when particle physicists tried to understand the observed spin zero pion scattering. The way to construct the non-linear realisation for general groups, used in this context, was given in reference [12]. 
The understanding  that pion scattering could be accounted for in this way was crucial for the development of particle physics in that it lead to the realisation  that symmetry was to play a central role. Until more recent times there were  very few  papers  which constructed a non-linear realisation involving  particles other than spin zero, but one of the most important  was the formulation of four dimensional gravity as the non-linear realisation the semi-direct product of $GL(4)$ and its vector representation [14,15]. Although  to arrive at Einstein's theory of general relativity in this way required imposing additional symmetries  beyond the symmetries of the non-linear realisation. 
\par
It was universally thought that the hidden symmetries found in the maximal supergravity theories in  lower dimensions were a consequence of the dimensional reduction procedure. However, in  2001 it was proposed  that there should exist a theory in eleven dimensions that had a vast symmetry group and this symmetry transformed all the fields of the supergravity theory and not just the scalar fields. It was found, in the original $E_{11}$ paper [13] that if one wanted to formulate the eleven dimensional supergravity theory as a non-linear realisation then one required a vast algebra, specifically a  rank eleven Kac-Moody called $E_{11}$. In 2003 it was understood how to include spacetime and it was proposed that the low energy effective theory of string and branes was given by 
{\bf E theory meaning  the non-linear realisation of the semi-direct product of $E_{11}$ and its vector representation, denoted $E_{11}\otimes_s l_1$} [16].
\par
The above one  sentence is a complete description of E theory but to work out its consequences has taken many years. E theory contains an infinite number of fields  in the adjoint representation of $E_{11}$ and these depend on an infinite dimensional spacetime whose coordinates are contained in the vector representation. It has been  shown that E theory contains, at low levels, the fields of all  the maximal supergravity theories and that these obey the equations of motion of these theories. The  supergravity theories in the different dimensions emerge by taking different decompositions of $E_{11}$ [17,18,19,20] and the maximal gauge supergravity theories appear if certain of the higher level fields are non-zero [21,22].  
\par
 In eleven dimensions one finds at low levels  the graviton, the three form, the six form, the dual graviton, ....[13] while at low levels the spacetime  contains the usual coordinate of spacetime $x^\mu$,   two form coordinates $x_{\mu\nu}$,  five form coordinates  $x_{\mu_1\ldots \mu_5}$, ...
  [16,23].  The equations of motion of the fields follow essentially uniquely from  E theory  and they have  been worked out at the non-linear level up to level three, that is, for the graviton, three-form, six-form [24,25] and the dual graviton [26]. If one takes the fields to only depend on the usual coordinates of spacetime then we find the equations of motion of eleven dimensional supergravity. The equations of motion have also been also worked out at the linearised level for the fields at level four [27] which provide an eleven dimensional origin for Romans theory [28]. 
  \par
 The field equations in E theory have been worked out at low levels  in  five [25] and seven [29] dimensions. If one restricts the generalised space-time to be just the usual space-time then the equations of motion of the low level fields also agree precisely with those of the corresponding  bosonic sectors of the supergravity  theories. For a review, see [30]. 
\par
There is no listing of the generators in $E_{11}$, or any other Kac-Moody algebra for that matter, and as a result there is no listing of the fields in E theory. However, the fields in E theory in eleven dimensions  that  have no blocks of ten or eleven indices are known [31] and they  are dual formulations of the usual fields of the supergravity fields. While some of the other fields are known to lead to the  gauge supergravities, the meaning of the remaining  fields is not known. 
 \par
The irreducible representation corresponding to the dynamics of the non-linear realisation has been worked out completely and it showed that the only dynamical degrees of freedom in eleven dimensions are those of the graviton and the three-form [32,33]. Thus although E theory contains an infinite set of dual fields, the only degrees of freedom are those of the maximal supergravity theories.  
 \par
The role of  the coordinates of the spacetime in E theory  is  less understood. These are encoded in the vector representation of $E_{11}$  which  contains all the brane charges [16,34,35,23]. Indeed for every brane charge, and so a brane, there is a coordinate. In eleven dimensions the low level coordinates are a follows 
$$
x^\mu, x_{\mu_1\mu_2} , x_{\mu_1\ldots \mu_5},  x_{\mu_1\ldots \mu_8},   x_{\mu_1\ldots \mu_7,b}, 
x_{\nu_1\nu_2\nu_3, \mu_1\ldots \mu_8}, x_{\nu_1\nu_2, \mu_1 \ldots \mu_{9}},  x_{(\nu_1\nu_2), \mu_1\ldots \mu_{9}}, x_{\nu,\mu_1\ldots \mu_{10}},\ldots 
\eqno(1.1)$$
\par
In lower dimensions the generators  of the vector representation  that have one block of anti-symmetrised indices are known. Not showing the familiar spacetime translations generators $P_a$ in the usual spacetime  they are  given  in the table below [34,35,36].  
The coordinates of the spacetime are in a one to one correspondence with the generators of  the vector representation and they belong to the same representation of the exceptional symmetries and so can be read of from the table. 
At level zero we have  the usual coordinates of the spacetime in $D$ dimensions that we are familiar with, but  at level one we find   coordinates which are scalars under the SL(D) transformations of our usual spacetime, and so also Lorentz transformations,  but belong to non-trivial representations of $E_{11-D}$. In particular, they belong to the 
 10,  {16},  {27},  56,   and  248+1 ,representations of  SL(5), SO(5,5), $E_6$,  $E_7$  and  $E_8$. 
 for $  D= 7,6,5,4$ and $3$    dimensions  respectively. 
\par
\vskip 1cm
 \eject
\medskip 
{\centerline{\bf {Table 2. The form generators  in the $l_1$ representation  in D
dimensions}}}
\medskip
$$\halign{\centerline{#} \cr
\vbox{\offinterlineskip
\halign{\strut \vrule \quad \hfil # \hfil\quad &\vrule Ê\quad \hfil #
\hfil\quad &\vrule \hfil # \hfil
&\vrule \hfil # \hfil Ê&\vrule \hfil # \hfil &\vrule \hfil # \hfil &
\vrule \hfil # \hfil &\vrule \hfil # \hfil &\vrule \hfil # \hfil &
\vrule \hfil # \hfil &\vrule#
\cr
\noalign{\hrule}
D&G&$Z$&$Z^{a}$&$Z^{a_1a_2}$&$Z^{a_1\ldots a_{3}}$&$Z^{a_1\ldots a_
{4}}$&$Z^{a_1\ldots a_{5}}$&$Z^{a_1\ldots a_6}$&$Z^{a_1\ldots a_7}$&\cr
\noalign{\hrule}
8&$SL(3)\otimes SL(2)$&$\bf (3,2)$&$\bf (\bar 3,1)$&$\bf (1,2)$&$\bf
(3,1)$&$\bf (\bar 3,2)$&$\bf (1,3)$&$\bf (3,2)$&$\bf (6,1)$&\cr
&&&&&&&$\bf (8,1)$&$\bf (6,2)$&$\bf (18,1)$&\cr Ê&&&&&&&$\bf (1,1)$&&$
\bf
(3,1)$&\cr Ê&&&&&&&&&$\bf (6,1)$&\cr
&&&&&&&&&$\bf (3,3)$&\cr
\noalign{\hrule}
7&$SL(5)$&$\bf 10$&$\bf\bar 5$&$\bf 5$&$\bf \overline {10}$&$\bf 24$&$\bf
40$&$\bf 70$&-&\cr Ê&&&&&&$\bf 1$&$\bf 15$&$\bf 50$&-&\cr
&&&&&&&$\bf 10$&$\bf 45$&-&\cr
&&&&&&&&$\bf 5$&-&\cr
\noalign{\hrule}
6&$SO(5,5)$&$\bf \overline {16}$&$\bf 10$&$\bf 16$&$\bf 45$&$\bf \overline
{144}$&$\bf 320$&-&-&\cr &&&&&$\bf 1$&$\bf 16$&$\bf 126$&-&-&\cr
&&&&&&&$\bf 120$&-&-&\cr
\noalign{\hrule}
5&$E_6$&$\bf\overline { 27}$&$\bf 27$&$\bf 78$&$\bf \overline {351}$&$\bf
1728$&-&-&-&\cr Ê&&&&$\bf 1$&$\bf \overline {27}$&$\bf 351$&-&-&-&\cr
&&&&&&$\bf 27$&-&-&-&\cr
\noalign{\hrule}
4&$E_7$&$\bf 56$&$\bf 133$&$\bf 912$&$\bf 8645$&-&-&-&-&\cr
&&&$\bf 1$&$\bf 56$&$\bf 1539$&-&-&-&-&\cr
&&&&&$\bf 133$&-&-&-&-&\cr
&&&&&$\bf 1$&-&-&-&-&\cr
\noalign{\hrule}
3&$E_8$&$\bf 248$&$\bf 3875$&$\bf 147250$&-&-&-&-&-&\cr
&&$\bf1$&$\bf248$&$\bf 30380$&-&-&-&-&-&\cr
&&&$\bf 1$&$\bf 3875$&-&-&-&-&-&\cr
&&&&$\bf 248$&-&-&-&-&-&\cr
&&&&$\bf 1$&-&-&-&-&-&\cr
\noalign{\hrule}
}}\cr}$$
\medskip

 The coordinates at higher levels than those of the usual  spacetime play a crucial role in the construction of the dynamics in E theory. However, to recover the supergravity theories we take the fields to only depend on the usual coordinates of spacetime. This 
 breaks the $E_{11}$ symmetry, but as explained in reference [37], it just corresponds to the fact that one is constructing a theory that contains only  point particles and are not taking account of the presence of branes. Indeed the $E_{11}$ symmetry transforms point particles into branes and so if one just considers the former it is inevitable that one will break the $E_{11}$ symmetry. We refer the reader to reference [37] for more details. 
 \par
 One of the most interesting aspect of E theory is the presence of its very extended spacetime which corresponds to recording positions in spacetime using objects other than point particles and in particular branes. To better  understand the role of the extra coordinates it is necessary to understand how they appear in E theory and this is the subject  of this paper. 
 \par
Inspired by T duality in  string theory a point particle field theory which was invariant under SO(10,10) was constructed [38,39]. 
 It has been referred to as  Siegel theory, but it is also called Doubled Field Theory. This theory  contained the graviton, the  two form and the dilaton of the massless sector of NS-NS sector of the superstring.  These fields depended on a spacetime with the coordinates $z^\Pi=(x^\mu ,y_{\dot \mu})$ that had twenty dimensions transforming in the vector representation of SO(10,10). The action was invariant under gauge transformations with a parameter that was also in the vector representation of SO(10,10) but only if one restricted the fields to obey certain  conditions. These were later  called section conditions and they are  of two types  
 $$
 \partial_\Pi \partial^\Pi \bullet=0 ,\ \rm {weak\ section \ condition} 
 \eqno(1.2)$$
 or 
$$ \partial_\Pi \star \partial^\Pi \bullet=0 ,\ \rm{strong \  section \ condition} 
 \eqno(1.3)$$
 where $\bullet$ and $\star$ were any two fields, parameters or any other quantities. In these equations  $\partial_\Pi= {\partial\over \partial z^\Pi}$, $\partial^\Pi= \Omega^{\Pi\Sigma} \partial _{\Sigma}$ and $ \Omega^{\Pi\Sigma}$ is  the SO(10,10) metric, see appendix B. These  conditions were motivated by the level matching condition in string theory. Since they were introduced in reference [39] the conditions have been systematically used in all subsequent papers on Siegel theory and in other constructions which used analogous conditions. Such generic equations of the above type are not usual in physics, at least in the experience of this author, and one can wonder if there is another way to proceed and replace them with more natural equations. To find such alternative conditions  is the task  undertaken in this paper. 
 \par
 In  reference [40] Double Feld Theory was derived from the toroidially compactified bosonic string and further developed [41] by setting it in a  geometrical setting. A clear and detailed account  of the gauge invariance of the actions using the above section conditions was given in references [43]. The relation to string field theory was also studied in [44]. The detailed correspondence with the formalism of Siegel theory [38,39] was given in [45]. 
 \par
 The connection between Siegel theory and E theory was was not immediately apparent but it was shown in  reference [46] that it appeared at level zero in  the non-linear realisation of $E_{11}\otimes_s l_1$ in the decomposition that leads to the IIA theory. 
 The extension of Siegel theory to include the massless  fields of the Ramond-Ramond was  found by evaluating E theory in its IIA decomposition at level one [47].  A derivation of this  same result from the view point of Siegel theory was later given in references  [48,49]. 
 \par
The way results were derived in E theory and Siegel theory was different.  In Siegel theory  the dynamics is  essentially determined  from local transformations which  include general coordinate and gauge transformations, subject to the section conditions. In E theory they are derived using the symmetries of the non-linear realisation which are the rigid $E_{11}$ symmetries  and the local tangent space symmetry. That the results in E theory are    essentially unique is a consequence of the fact that $E_{11}$ is a duality symmetry leading to  an infinite number of duality relations. The resulting equations of motion have been found  to be  invariant under the usual gauge and general coordinate symmetries even though this was not required from the outset. In this method of derivations no use was made of section conditions or any similar conditions. 
\par
This was true  for all the theories  contained  in E theory with the exception of the ten dimensional  IIA theory at level zero. In this case the dynamics  did not follow uniquely from the symmetries of the non-linear realisation at level zero. This was due to the fact that the  the tangent space group,  $I_c(E_{11})$ at level zero,   is only $SO(10)\otimes SO(10)$.  However, carrying out the calculation of the equations of motion at higher levels the dynamics would follow uniquely  as it does in eleven dimensions. This would require the calculation up  to level two where the duals of the level zero fields appear. 
\par
 In section two we will review the parts of E theory that we will need in this paper. In section three we study the local transformations previously  proposed [50], but have not been widely used in E theory. We show that they contain the well known  general coordinate and gauge transformations. We also show that they have a closing algebra provided one adopts a condition on the parameters that restricts their dependence of the enlarged spacetime of E theory, namely equation (3.5). 
 In section four we discuss the irreducible representation of E theory  that corresponds to the supergravity theories [32,33] and in particular explain how  the   dependence of the fields on the coordinates of the spacetime that are beyond those of our usual spacetime arises from the presence of branes. We also explain the need for the fields to satisfy new equations,  in addition to their usual equations of motion,   that constrains the dependence of the fields on these additional coordinates. 
 \par
 In section five we give a brief review of  E theory in its decomposition at level zero that leads to the IIA theory in ten dimensions [46] which has  the same fields and spacetime as Siegel theory (Double field theory)[38,39]. We show that the local transformations of E theory [50], at level zero in the IIA decomposition, agree with those found previously in this theory. However, the closure condition adopted in section three is not one of  the section conditions of equations (1.2) and (1.3). Finally we discuss what conditions for the action of Siegel theory to be invariant under the local transformations. While we recover the well known fact that this is true if one assumes the section conditions we argue that it would be more natural to take the fields to satisfy certain specific equations,  in addition to their equations of motion,  which restrict their dependence on the additional coordinates of spacetime.

\medskip
{\bf 2. A very brief review of E theory.} 
\medskip
In this section we give  a very short summary of the properties of the non-linear realisation of the semi-direct product of the $E_{11}$ and the vector representation, denoted by $E_{11}\otimes_s l_1$,  which we will need in this paper.  The reader can find  detailed accounts  in the earlier papers on $E_{11}$ and in particular in the papers  [24,25] and the review [30]. In order to be concrete,   we will explain E theory in the context of eleven dimensions. We denote the generators of $E_{11}$ by $R^\alpha$ and those of the vector representation by $l_A$. The algebra  $E_{11}\otimes_s l_1$ is of the form 
$$
[ R^\alpha , R^\beta]= f^{\alpha\beta}{}_{\delta} R^\delta, \ [ R^\alpha , l_A ] = -(D^\alpha)_A{}^B l_B , \ [l_A, l_B]=0
\eqno(2.1)$$
where the generators of $E_{11}$,  from the view point of eleven dimensions,  are 
$$
R^\alpha =\{ \ldots  ,\  R_{a_1a_2\ldots a_8,b} , \  R_{a_1a_2\dots a_6} , \ R_{a_1a_2a_3} ,  K^a{}_b , \ R^{a_1a_2a_3} , \ R^{a_1a_2\dots a_6} , \   R^{a_1a_2\ldots a_8,b}, \ldots \}
\eqno(2.2)$$
and those of the vector representation are 
$$
l_A=\{P_a, Z^{ab}, \ Z^{a_1\ldots a_5}, \ Z^{a_1\ldots a_7,b},\  Z^{a_1\ldots a_8},\ 
\  Z^{ b_1 b_2  b_3,  a_1\ldots  a_8},\ \ldots \}
\eqno(2.3)$$
 The indices range over $a,b,\ldots =1,2\ldots 11$.  The generators are labelled by a level. At level zero they are the  $K^a{}_b$ which generate  GL(11).  
\par
The lowest level commutators are 
$$
[K^a{}_b , K^c{}_d ] = \delta ^b_c K^a{}_d - \delta ^a_d K^b{}_c ,\ [K^a{}_b , R^{c_1c_2c_3} ] = 3\delta _b^{[ c_1} R^{| a |  c_2c_3 },\ 
$$
$$
[ R^{a_1a_2a_3}, R^{b_1b_2b_3} ] = 2R^{a_1a_2a_3b_1b_2b_3} , 
$$
$$
[ R^{a_1\ldots a_3}, R_{b_1\ldots b_3}]= 18 \delta^{[a_1a_2}_{[b_1b_2}
K^{a_3]}{}_{b_3]}-2\delta^{a_1a_2 a_3}_{b_1b_2 b_3} \sum_c K^c{}_c 
\eqno(2.4)$$
\par
The commutators of the generators of $E_{11}$ with those of the vector representation at low levels are given by 
$$
[K^a{}_b , P_c ]= - \delta ^a_c P_b +{1\over 2} \delta^a_b P_c ,\ 
[R^{a_1a_2a_3} , P_b ]= 3\delta _b^{[a_1} Z^{a_2a_3]},
$$
$$ 
 [R_{a_1a_2a_3} , Z^{b_1b_2} ]=6 \delta^{b_1b_2} _{[ a_1a_2} P_{a_3 ]}
\eqno(2.5)$$
The reader can find the $E_{11}\otimes_s l_1$, algebra up to level three in chapter 16 of reference [19] or, for example,  in appendix A of reference [46]. 
\par
The Killing form $C^{\alpha\beta}$ of $E_{11}$ is 
$$
C^a{}_{b , }{}^c{}_d\equiv ( K^a{}_b ,  K^c{}_d)=\delta^a_d\delta^c_b-{1\over 2} \delta^a_b\delta ^c_d ,\ 
C^{a_1a_2}{}_{,b_1b_2b_3}\equiv (R^{a_1a_2a_3}, R_{b_1b_2b_3})= 6\delta^{a_1a_2a_3}_{b_1b_2b_3}, \ 
$$
$$
C^{a_1\ldots a_6}{}_{,b_1\ldots b_6}\equiv (R^{a_1\ldots a_6}, R_{,b_1\ldots b_6})= -180\delta^{a_1\ldots a_6}_{b_1\ldots b_6}, \ldots 
\eqno(2.6)$$
where all  other components involving these generators vanish.  
The inverse Killing form  $C^{-1}$ is given by 
$$
(C^{-1})^a{}_{b , }{}^c{}_d =\delta^a_d\delta^c_b-{1\over 9} \delta^a_b\delta ^c_d ,\ 
(C^{-1})^{a_1a_2a_3}{}_{,b_1b_2b_3}= {1\over 6}\delta^{a_1a_2a_3}_{b_1b_2b_3} , \ 
C^{-1}{}^{a_1\ldots a_6}{}_{,b_1\ldots b_6} = -{1\over 180}\delta^{a_1\ldots a_6}_{b_1\ldots b_6},
\eqno(2.7)$$
\par
The nonlinear realisation  is constructed from  a group element $g$ of $E_{11}\otimes_s l_1$ so that it invariant under the transformations
$$
g\to g_0g h,\quad {\rm for } \quad g_0\in E_{11} , \ h \in I_c(E_{11})
\eqno(2.8)$$
where $g_0$ is a transformation which is independent of spacetime while $h$ depends on the spacetime and belongs to $ I_c(E_{11})$,  the Cartan involution subalgebra of $E_{11}$. 
\par
The group element $g$ can be written in the form $g=g_l g_E$ where 
$$
g_E=  \ldots e^{ h_{a_1\ldots a_{8},b}
R^{a_1\ldots a_{8},b}} e^{ A_{a_1\ldots
a_6} R^{a_1\ldots a_6}}e^{ A_{a_1\ldots a_3} R^{a_1\ldots 
a_3}} e^{h_a{}^b K^a{}_b} \equiv e^{A_{\underline \alpha} R^{\underline \alpha}}
\eqno(2.9)$$ 
and $$
g_l= e^{x^aP_a} e^{x_{ab}Z^{ab}} e^{x_{a_1\ldots a_5}Z^{a_1\ldots a_5}}\ldots \equiv
e^{x^A l_A} 
\eqno(2.10)$$
We have used the local subalgebra $I_c(E_{11})$  to choose the group element $g_E$ to have no negative level generators. The $h_a{}^b$, $A_{a_1a_2a_3}, \ldots $ will be the fields of the theory and they depend on the coordinates $x^a$, $x_{a_2a_2}, \ldots $
\par
To construct the dynamics we consider the Cartan forms which are given by 
$$
{\cal V}\equiv g^{-1} d g=  dz^\Pi E_\Pi{}^A l_A +  dz^\Pi G_{\Pi, \underline \alpha} R^{\underline \alpha}=
 g_E^{-1} dz^A l_A g_E +g_E^{-1} dg_E \equiv {\cal V}_A+ {\cal V}_l
\eqno(2.11)$$
The $G_{\Pi, \underline \alpha} $ are the Cartan forms of $E_{11}$ while $E_\Pi{}^A$  is the  vierbein  on the spacetime. 
Using equation (2.1) we find that  $ k^{-1} l_\Pi k= D(k)_\Pi{}^\Sigma l_\Sigma$  for $k\in E_{11}$  and $D(k)_\Pi{}^\Sigma$   is the  
matrix representative in the vector representation. As a result   we identify  $E_\Pi{}^A= D(g_E)_\Pi{}^A$. 
\par
The Cartan forms  only transform under the local transformations  as follows 
$$ 
{\cal V}\to h^{-1}{\cal V} h + h^{-1} d h\quad  {\rm or \ equivalently }\ {\cal V}_A\to h^{-1}{\cal V}_A h + h^{-1} d h\quad {\rm and }\quad 
{\cal V}_l\to h^{-1}{\cal V}_l h 
\eqno(2.12)$$
The coordinates $z^\Pi$ are inert under $I_c(E_{11})$ transformations but under a  rigid $g_0\in E_{11}$ they transformation as 
$$
 dz^{\Pi \prime }= dz^{ \Sigma}D(g_0^{-1})_\Sigma{}^\Pi ,\ \rm{ and }\   \partial_\Pi^\prime\equiv {\partial\over \partial z^{\Pi\prime}}=  D(g_0)_\Pi{}^\Sigma\partial_\Sigma
\eqno(2.13)$$ 
The vierbein  transforms under  local $I_c(E_{11})$ tangent group transformations on its tangent index and rigid $E_{11}$ transformations on its world index. More precisely under a rigid $g_0\in E_{11}$ transformation the vierbein transforms as 
$$E_\Pi{}^{A\prime}= D(g_0)_\Pi{}^\Sigma E_\Sigma{}^A
\eqno(2.14)$$
While under the local $I_c(E_{11})$ transformation   $h=( R^\alpha- R^{-\alpha})_B{}^A \Lambda_{\alpha}$   
the  vierbein transforms  as 
$$
 \delta E_\Pi{}^A= E_\Pi{}^B(D^\alpha- D^{-\alpha})_B{}^A \Lambda_{\alpha}
 \eqno(2.15)$$
\par
Using the group element of equations (2.9) and (2.10) we can  find that the vierbein and up to level three  it  is given by  [51,52]
$$
{ E}= (\det e)^{-{1\over 2}}
\left(\matrix {e_\mu{}^a&-3 e_\mu{}^c A_{cb_1b_2}& 3 e_\mu{}^c A_{cb_1\ldots b_5}+{3\over 2} e_\mu{}^c A_{[b_1b_2b_3}A_{|c|b_4b_5]}\cr
0&(e^{-1})_{[b_1}{}^{\mu_1} (e^{-1})_{b_2]}{}^{\mu_2}&- A_{[b_1b_2b_3 }(e^{-1})_{b_4}{}^{\mu_1} (e^{-1})_{b_5 ]}{}^{\mu_2}  \cr
0&0& (e^{-1})_{[b_1}{}^{\mu_1} \ldots (e^{-1})_{b_5]}{}^{\mu_5}\cr}\right)
\eqno(2.16)$$
\par
Using the form of the group element of equation (2.9),  the Cartan forms can  be found to be given, up to level three,  by
$$
G _{a}{}^b=(e^{-1}d e)_a{}^b,\ \ G_{a_1\ldots a_3}= e_{a_1}{}^{\mu_1}\ldots e_{a_3}{}^{\mu_3}
dA_{\mu_1\ldots \mu_3}, 
$$
$$ 
G_{a_1\ldots a_6}=  e_{a_1}{}^{\mu_1}\ldots e_{a_6}{}^{\mu_6}(d A_{\mu_1\ldots \mu_6} 
- A_{[ \mu_1\ldots \mu_3}d A_{\mu_4\ldots \mu_6]})
\eqno(2.17)$$
 \par
 The very early papers on $E_{11}$ constructed,  at low levels,  the non-linear realisation of $E_{11}$  taking only the usual spacetime [13, 53]. In [16] it was proposed to take the non-linear realisation of  $E_{11}\otimes_sl_1$, so introducing the coordinates of spacetime which belonged to the vector ($l_1$) representation [16]. This non-linear realisation was constructed in part at low levels, but including  additional coordinates to those of the usual spacetime  in the following papers: 
 \item {-}    in five dimensions in the context of gauged supergravities  [20], 
  \item {-}   parts of  four dimensional maximal supergravity using a viebein associated with the additional level one coordinate, but also using  diffeomorphism invariance to constrain the dynamics  [54,55], 
  \item {-}   construction of the IIA theory [46,47], in  ten dimensions
  \item{-}  construction of   the duality relations in eleven dimensions  [51]  
  \item {-}  construction of  duality relations   in four dimensions [56]. 
    \item {-}  a  systematic construction of the low level dynamics of the non-linear realisation was subsequently given in eleven [24,25,26,27],   five dimensions [24] and seven dimensions [29].

\medskip
{\bf  3. Local  transformations in E theory}
\medskip
The local transformation of the vierbein  in the $E_{11}$ non-linear realisation was proposed to be given by [50]
$$
\delta E_\Pi{}^A= (C^{-1})_{\alpha\beta} (D^\alpha)_{\Pi}{}^\Sigma E_\Sigma{}^A (D^\beta)_\Lambda{}^\Gamma \partial_\Gamma \Lambda^\Lambda + \Lambda^\Pi \partial_\Pi E_\Pi{}^A
\eqno(3.1)$$
where the sum runs over the roots $\alpha$ of $E_{11}$,  $C^{-1}$ is the inverse Killing matrix of $E_{11}$ and the matrix $(D^\alpha)_{\Pi}{}^\Sigma$ is the matrix representative of $E_{11}$ in the vector representation, see equation (2.1). Since the vierbein contains all the fields of the theory equation (3.1) also gives  their transformations. 
\par
The vierbein also transforms under the  Cartan involution invariant subalgebra $I_c(E_{11})$ as given in equation (2.8). As explained in section two, one generally makes a choice of group element using this local symmetry  and as a result  the expression for the vierbein of equation (2.16) is not the most general one. As such when carrying out procedures on the vierbein, such as the gauge transformation of equation (3.1),   one usually  has to make a compensating local $I_c(E_{11})$ to maintain its form. 
\par
Carrying out the commutators of two gauge transformations we find ,after a straight forward calculation,  that 
$$
[\delta_{\Lambda_1} , \delta_{\Lambda_2} ] E_\Pi{}^A= \Lambda_c ^\Sigma \partial_\Sigma E_\Pi{}^A+
(C^{-1})_{\alpha\beta} (D^\alpha)_{\Pi}{}^\Sigma E_\Sigma{}^A (D^\beta)_\Lambda{}^\Gamma \partial_\Gamma \Lambda_c^\Lambda
+ (D^\alpha)_{\Pi}{}^\Sigma E_\Sigma{}^A C_\alpha
\eqno(3.2)$$
where 
$$
\Lambda_c^\Gamma= \Lambda_2^\Sigma \partial_\Sigma \Lambda_1^\Gamma-\Lambda_1^\Sigma \partial_\Sigma \Lambda_2^\Gamma
\eqno(3.3)$$
and 
$$
C_\alpha= f^{\gamma\delta}{}_\alpha (D_\gamma)_{\Pi}{}^\Sigma \partial_\Sigma \Lambda_1 ^\Pi (D_\delta)_{\Gamma}{}^\Theta \partial_\Theta \Lambda_2 ^\Gamma - (D_\alpha)_{\Pi}{}^\Sigma [\partial_\Sigma  \Lambda_2^\Gamma \partial_\Gamma \Lambda_1^\Pi-\partial_\Sigma  \Lambda_1^\Gamma \partial_\Gamma \Lambda_2^\Pi]
\eqno(3.4)$$
Here $f^{\gamma\delta}{}_\alpha $ is the structure constant of the $E_{11}$ algebra. Clearly if we were to set $C_\alpha=0$ the gauge algebra would close to give a commutator of a general  form familiar from general coordinate transformations. Thus we adopt this condition on the parameters of the gauge transformations and we list it as an equation for future reference 
 $$
C_\alpha= f^{\gamma\delta}{}_\alpha (D_\gamma)_{\Pi}{}^\Sigma \partial_\Sigma \Lambda_1 ^\Pi (D_\delta)_{\Gamma}{}^\Theta \partial_\Theta \Lambda_2 ^\Gamma - (D_\alpha)_{\Pi}{}^\Sigma [\partial_\Sigma  \Lambda_2^\Gamma \partial_\Gamma \Lambda_1^\Pi-\partial_\Sigma  \Lambda_1^\Gamma \partial_\Gamma \Lambda_2^\Pi]=0
\eqno(3.5)$$
\par
We will now gather further evidence that the gauge transformation of equation (3.1) is indeed correct which in turn implies the condition of equation (3.5). We can derive the gauge transformations using the Noether method  which is usually used in the context of deriving an action but it can also be used to find the local transformations, as was done for the IIB supergravity theory [9]. For a review of the Noether method,  see section 13.1.1 of reference [19]. The advantage of this approach is that it makes almost no assumptions except those which are essentially obvious. 
\par
The group element in the $E_{11}$ non-linear realisation is of the generic  form $g=e^{A_\alpha R^\alpha}$.  As we are beginning with the linearised theory we start with the fields $A_\alpha $,  or equivalently  $ A_\Pi{}^A\equiv A_\alpha (D^\alpha)_\Pi{}^A$. The distinction between world and tangent indices does not exist at the linearised level. 
The gauge transformations of $A_\alpha $ must be linear in the parameter $\Lambda^\Gamma$ and the derivatives $\partial_\Pi$ which are in the $\bar l_1$ (dual vector)   and $l_1$ (vector) representations respectively. Demanding that the result be in the adjoint representation as it is the variation of $A_\alpha $, it was shown that the linearised gauge transformation had to be of the form [50]
$$
\delta A_\alpha =(D_\alpha)_\Pi{}^\Sigma \partial_\Sigma \Lambda^\Pi = (C^{-1})_{\alpha\beta} (D^\beta)_\Pi{}^\Sigma \partial_\Sigma \Lambda^\Pi \equiv N_\alpha \ \rm{or  } \ 
$$
$$
\delta A_\Xi{}^A= (C^{-1})_{\alpha\beta} (D^\alpha)_\Xi{}^A  (D^\beta)_\Pi{}^\Sigma \partial_\Sigma \Lambda^\Pi =(D^\alpha)_\Xi{}^A  N_\alpha \
\eqno(3.6)$$
The linearised theory should also   be invariant under rigid translations in the spacetime, that is, 
$$
\delta A_\alpha =\Xi^\Pi \partial_\Pi A_\alpha
\eqno(3.7)$$
where $\Xi^\Pi $ is a constant. 
\par
To find the non-linear theory, including its transformations, we let the rigid translations depend on the spacetime, $\Xi\to \Xi(x)$. One can only then find an invariant theory with  a closing algebra if we identify the parameter of the translations with that of the gauge transformations, that is,  $ \Xi^\Gamma= \Lambda^\Gamma$. Then the commutator of a now local translation with a local gauge transformation can be written as 
$$
[\delta_1 , \delta_2] A_\alpha= \Lambda_2^\Gamma\partial_\Gamma  (D_\alpha)_\Pi{}^\Sigma \partial_\Sigma \Lambda_1^\Pi -
\Lambda_1^\Gamma\partial_\Gamma  (D_\alpha)_\Pi{}^\Sigma \partial_\Sigma \Lambda_2^\Pi
$$
$$
= (D_\alpha)_\Pi{}^\Sigma \partial_\Sigma \Lambda_c^\Pi-  (D_\alpha)_\Pi{}^\Sigma \partial_\Sigma \Lambda_2^\Gamma \partial_\Gamma\Lambda_1^\Pi+ (D_\alpha)_\Pi{}^\Sigma \partial_\Sigma \Lambda_1^\Gamma \partial_\Gamma\Lambda_2^\Pi
\eqno(3.8)$$
where 
$ \Lambda_c^\Pi=  \Lambda_2^\Gamma\partial_\Gamma\Lambda_1 ^\Pi-  \Lambda_1^\Gamma\partial_\Gamma\Lambda_2 ^\Pi$.
While the first term is indeed a gauge transformation the second term is not. We might hope to cancel this term using a local gauge transformation but it is not obviously of the form of $N_\alpha$. To cancel this term it must be reformulated so as to contain both $N_{1\alpha}$ and $N_{2\alpha}$ and so it must be equal to 
$$
(D_\alpha)_\Pi{}^\Sigma \partial_\Sigma \Lambda_2^\Gamma \partial_\Gamma\Lambda_1^\Pi- (1\leftrightarrow 2)
= N_{1\gamma}N_{2\delta} f^{\gamma\delta}{}_\alpha
\eqno(3.9)$$
where $f^{\gamma\delta}{}_\alpha=- f^{\delta\gamma}{}_\alpha$ is a constant. Working with  $A_\Xi{}^A$ instead of $A_\alpha$ and taking $f^{\gamma\delta}{}_\alpha$ to be  the structure constant of $E_{11}$ we find that 
$$
[\delta_1,  \delta_2 ]  A_\Xi^A= (D^\alpha)_\Xi{}^A  (D_\alpha)_\Pi{}^\Sigma \partial_\Sigma \Lambda_c^\Pi 
- ((D^\gamma) _\Xi{}^ \Lambda (D^\delta) _ \Lambda {} ^A N_{1\gamma}N_{2\delta}- (1\leftrightarrow 2 ))
\eqno(3.10)$$
\par
We can now cancel the additional terms in equation (3.8) by adding a term involving the gauge field to its transformation. The result can be written as  
$$
\delta (\delta_\Xi^A+A_\Xi^A)= \Lambda^\Pi \partial_\Pi  (\delta_\Xi^A+A_\Xi^A)+(D^\alpha)_\Xi{}^\Pi (\delta_\Pi{}^A+A_\Pi{}^A)
  (D_\alpha)_\Gamma{}^\Sigma \partial_\Sigma \Lambda^\Gamma 
  \eqno(3.11)$$
This agrees with the gauge  transformation of equation (3.1) up to first order in the field. To see this we note that the vierbein  has the form $E_\Pi{}^A= (e^{A_\alpha D^\alpha})_\Pi{}^A$ and so at lowest order it is given by $E_\Pi{}^A=\delta _\Pi^A+ A_\alpha (D^\alpha)_\Pi{}^A$. This provides  strong evidence for the gauge transformation as we have made almost no assumptions. 
\par
 It would be interesting to carry out the Noether procedure at higher levels to confirm the gauge transformation to all order in the field and so the condition of equation (3.5). There is every reason to think this will work. 
  \par
 From this calculation it is clear why the condition of equation (3.5) is required. In the Noether procedure it is usually the case that the transformation of the gauge field is equal to a derivative acting directly on the parameter in the generic form 
$\delta A_\mu= \partial_\mu \Lambda$. However in the above case we have a projection involving the matrix $(D^\alpha)_\Pi{}^\Lambda$ that is not present in the spacetime translation and so one cannot cancel the terms that arise without adopting the above  constraint. 
\par
The gauge transformations of equation (3.1) do not look like the ones we are used to.  We will now evaluate this transformation  explicitly at low levels to get a better understand of what they contain. 
Let us first take the gauge parameter to be at level zero and so  of the form $\xi^\nu $ and also only take the derivative to  be  at level zero, that is, the usual derivative, $\partial_\mu$ and so we have  $\partial_\mu \xi^\nu$. Then in the gauge transformation of equation (3.1) we find the object 
$N_\alpha= (D_\alpha) _\nu{}^\mu\partial_\mu\xi^\nu$, but looking at the commutators of equation (2.5), we see that this can only be none zero for an $\alpha$ corresponding to the generator $K^a{}_b$ and so  the commutator 
$$
[ K^a{}_b , P_c]= -(D^a{}_b)_c{}^d P_d=- \delta_c^a P_b+{1\over 2} \delta _b^a P_c \ \rm { and \ so   } \ (D^a{}_b)_c{}^d
= \delta_c^a\delta^d_b- {1\over 2} \delta_b^a\delta ^d_c
\eqno(3.12)$$
Using this relation and the inverse Cartan matrix of equation (2.7)  we find that the transformation of the vierbein is given by 
$$
\delta E_\mu{}^A= (D^\rho{}_\kappa)_\mu{}^\lambda E_\lambda {}^A\partial_\rho \xi^\kappa+ \xi^\kappa \partial_\kappa E_\mu{}^A= \partial_\mu \xi^\lambda E_\lambda{}^A - {1\over 2}  \partial_\lambda\xi^\lambda  E_\mu{}^A+ \xi^\kappa \partial_\kappa E_\mu{}^A
\eqno(3.13)$$
\par
Looking at equation (2.16) we see that the component $E_\mu{}^a=(\det e)^{-{1\over 2}} e_\mu{}^a$ while 
$E_{\mu b_1b_2}=-3 (\det e)^{-{1\over 2}}  e_\mu^c A_{ab_1b_2}$ and so we find that 
$$
\delta e_\mu{}^a= \partial_\mu \xi^\lambda e_\lambda{}^a+ \xi^\kappa \partial_\kappa e_\mu{}^a, 
\ \delta A_{a_1a_2a_3} =  \xi^\kappa \partial_\kappa  A_{a_1a_2a_3} , \ldots 
\eqno(3.14)$$
which are indeed the transformations of a standard diffeomorphism. Thus the gauge transformation of equation (3.1) when restricted to contain just a usual derivative at level zero and the level zero parameter is just the well known diffeomorphism. 
\par
Proceeding in a similar  way we find that if we take the parameter to be at level one, that is, $\Lambda_{\mu_1\mu_2}$ and the derivative at level zero then 
$$
\delta E_\Pi {}^A= - (D^{\nu_1\nu_2\nu_3} ) _\Pi{}^\Sigma E_\Sigma {}^A    \partial_{\nu_1}\Lambda_{\nu_2\nu_3} 
\eqno(3.15)$$
As a result,  we conclude that 
$$\delta E_{\mu a_1a_2} =3 E^{\lambda_1\lambda_2}{}_{a_1a_2} \partial_{[\mu}\Lambda_{\lambda_1\lambda_2]}\ {\rm and \ so} \ 
\delta A_{\mu_1\mu_2\mu_3}= - \partial _{[\mu_1} \Lambda_{\mu_2\mu_3]}
\eqno(3.16)$$
which is the expected gauge transformation of the three form. 
\par
The gauge transformations of the vierbein involving  spacetime derivatives and parameters at level zero and one is given by 
$$
\delta E_\Pi{}^A= \xi^\mu\partial_\mu  E_\Pi{}^A+ \Lambda_{\mu\nu}\partial^{ \mu\nu} E_\Pi{}^A +(D^\mu{}_\nu)_\Pi{}^\Sigma E_\Sigma {}^A\partial_\mu \xi^\nu
-(D^{\nu_1\nu_2\nu_3})_\Pi{}^\Sigma E_\Sigma {}^A\partial_{\nu_1}\Lambda_{\nu_2\nu_3}
$$
$$
-{1\over 2} (D_{\nu_1\nu_2\nu_3})_\Pi{}^\Sigma E_\Sigma {}^A\partial^{\nu_1 \nu_2}\xi^{\nu_3}
-2(D^\rho{}_\kappa)_\Pi{}^\Sigma E_\Sigma {}^A\partial^{\kappa\tau}\Lambda_{\rho\tau}
+{1\over 3}(D^\rho{}_\rho)_\Pi{}^\Sigma E_\Sigma {}^A\partial^{\kappa\tau}\Lambda_{\kappa\tau}
\eqno(3.17)$$
\par
Using equation (3.17 ) we find that the gauge transformation involving just the parameters $\xi^\mu$ and $\Lambda_{\mu_1\mu_2}$ and the derivative $\partial^{\kappa_1\kappa_2}$ are given by 

$$
\delta E_\mu{}^A= -2 \partial^{\kappa \nu} \Lambda _{\mu\nu} E_{\kappa}{}^A - {1\over 2}\partial^{\kappa_1\kappa_2}  \Lambda_{\kappa_1\kappa_2}   E_\mu{}^A + \Lambda_{\kappa_1\kappa_2} \partial^{\kappa_1\kappa_2}   E_\mu{}^A 
\eqno(3.18)$$
and 
$$
\delta E^{\lambda_1\lambda_2 A}= 4\partial^{ [\lambda_1|  \nu} \Lambda_{\tau \nu} E^{\tau | \lambda_2]A
}-{3\over 2} \partial^{\kappa_1\kappa_2}  \Lambda_{\kappa_1\kappa_2}  E^{\lambda_1\lambda_2 A}
+ \Lambda_{\kappa_1\kappa_2} \partial^{\kappa_1\kappa_2}    E^{\lambda_1\lambda_2 A}
$$
$$
+3 \partial^{[\lambda_1\lambda_2 }\xi^{\tau]} E_{\tau}{}^{A}
\eqno(3.19)$$
\par
Taking $A=a$ in equation (3.19) we find that 
$$
\delta E^{\lambda_1\lambda_2 a}=3 \partial^{[\lambda_1\lambda_2 }\xi^{\tau]} E_{\tau}{}^{a}+\ldots 
\eqno(3.20)$$
where $+\ldots$ means terms which involve $E^{\lambda_1\lambda_2 a}$. However in our choice of group element of equation (2.9) the vierbein takes the form of equation (2.16) and so the gauge transformations do not preserve this chocie. To remedy matters we have to take a compensating $I_c(E_{11})$ transformation. Using equation (2.15) we find that such a transformation with parameter $\Lambda^{a_1a_2a_3}$ takes the form 
$$
\delta E_\Pi{}^a= 6 E_\Pi{}_{b_1b_2}\Lambda^{b_1b_2 a} ,\ \delta E_\Pi{}_{a_1a_2}= -3 E_\Pi{}^c\Lambda_{c a_1a_2}
+{5!\over 2}  E_\Pi{}_{a_1a_2c_1c_2c_3}\Lambda^{c_1c_2c_3}
\eqno(3.21)$$
To compensate for the gauge transformations of equation (3.20) and to restore the gauge choice $E^{\lambda_1\lambda_2 a}$ we choose 
$$
\Lambda^{a_1a_2a_3} = -{1\over 2}e_{\mu_1}^{a_1} e_{\mu_2}^{a_2} e_{\mu_3}^{a_3}  \partial ^{[\mu_1\mu_2}\xi^{\mu_3]}
\eqno(3.22)$$
As a result the transformation of $E_\mu{}^a$ is given by 
$$
\delta E_\mu {}^a= -2\partial^{\kappa \nu} \Lambda_{\mu\nu} E_\kappa{}^a -{1\over 2} \partial^{\kappa_1\kappa_2}  \Lambda_{\kappa_1\kappa_2} E_\mu{}^a 
+   \Lambda_{\kappa_1\kappa_2} \partial^{\kappa_1\kappa_2} E_\mu{}^a 
$$
$$
-{3\over 2} \partial^{[\tau_1\tau_2} \xi^{\tau_3 ] }e_{\tau_1}{}^{b_1} e_{\tau_2}{}^{b_2}e_{\tau_3}{}^{a} E_{\mu b_1b_2}
\eqno(3.23)$$
Using the values of the components of the vierbein of equation (2.16 ) we find that 
$$
\delta e_\mu{}^a= -2\partial^{\kappa \nu} \Lambda_{\mu\nu} e_\kappa{}^a +   \Lambda_{\kappa_1\kappa_2} \partial^{\kappa_1\kappa_2} e_\mu{}^a +{1\over 3} \partial^{\kappa_1\kappa_2}  \Lambda_{\kappa_1\kappa_2} e_\mu{}^a 
$$
$$
+{9\over 2} \partial^{[\tau_1\tau_2} \xi^{\tau_3 ] }e_{\tau_1}{}^a  A_{\mu\tau_2\tau_3} 
-{1\over 2}  \partial^{[\tau_1\tau_2} \xi^{\tau_3 ] }A_{\tau_1\tau_2\tau_3 }e_\mu{}^a
\eqno(3.24)$$
as well as the previously discussed general coordinate transformation. This agrees with the linearised result found in reference [50]. 
The gauge variations of the other fields can be calculated in a similar way.


\medskip
{\bf 4. Irreducible representations in E theory and extended spacetime}
\medskip
One place where we know precisely what is  the dependence of the fields on the spacetime of  E theory  is for  the irreducible representations of E theory, that is, the irreducible representations of $I_c(E_{11})\otimes_s l_1$ [32,33]. This is the E theory analogue of the representations of the Poincare group, indeed at level zero the algebra  $I_c(E_{11})\otimes_s l_1$  is just the Poincare algebra. The  vector representation contains all the branes charges which we can write as 
$$
p_a,\ p^{a_1a_2}, \ p^{a_1\ldots a_5}, \ldots 
\eqno(4.1)$$
\par
Following  the Wigner procedure we select  fixed values for the momenta and then compute the little group. The massless representation arising from choosing $p_{-}\equiv {1\over \sqrt{2} }(p_9-p_0)= \sqrt {2} m$,  $p_{+}\equiv {1\over \sqrt{2} }(p_9+p_0)=0=p_i$,  with  other momenta being zero, was constructed in reference [32]. The little algebra that preserves this choice of momenta  just  contains all generators of  $I_c(E_{11})$ which carry no lower $-$ indices. However all   such  generators with lower  $+$ indices all commute and, following the usual procedure, we take them to be trivially realised. Then we are left with generators with no lower  $+$ or $-$ indices and so arive at the algebra    $I_c(E_9)$, namely 
$$
I_c(E_9)=\{J_{ij} , \ S_{i_1i_2i_3} ,\ S_{i_1\ldots i_6} ,\  S_{i_1\ldots i_8,j } ,\  \ S_{j_1\ldots j_9, i_1i_2i_3} ,\ S_{j_1\ldots j_9, i_1\ldots i_6} ,\
S_{j_1\ldots j_9, i_1\ldots i_8,j }  ,\ \ldots \}
\eqno(4.2)$$
where the indices take the range $i, j , \ldots = 2,\ldots , 10$ and 
$$
J_{ij} = K^k{}_j\eta_{ki} - K^k{}_i\eta_{kj}, S_{i_1i_2i_3}= R^{j_1j_2j_3}\delta_{i_1j_1} \delta_{i_2 j_2} \delta_{i_3j_3} - R_{i_1i_2i_3}, \ \ldots 
\eqno(4.3)$$
\par
A representation of $I_c(E_9)$ is provided by the Cartan involution generators of $E_9$ which are have a minus sign under the action of the Cartan involution. We  label this representation by $\psi_I(0)$ and so 
$$
\psi_I(0)=\{ h_{ij}(0)= h_{(ji)}(0), \ A_{i_1i_2i_3}(0) ,\ A_{i_1\ldots i_6} (0),\  A_{i_1\ldots i_8,j }(0) ,\  \ A_{j_1\ldots j_8, i_1i_2i_3} (0), 
$$
$$
A_{j_1\ldots j_9, i_1\ldots i_6} (0),\ \ldots \} ,\ \  
\eqno(4.4)$$
However, all these fields are related to $h_{ij}(0)$ and $A_{i_1i_2i_3}(0)$ by $I_c(E_9)$ covariant duality relations, and so the theory only contains the degrees of freedom associated with  the graviton and the three form. 
\par
The generators of $I_c(E_{11})$ which are   outside the little group are 
$$
J_{-i} , \ J_{+-},\  S_{+-i},\ \ S_{-i_1i_2} ,\ S_{-i_1\ldots i_5} ,\  S_{-+i_1\ldots i_4} , \ldots 
\eqno(4.5)$$
The general state in the representation is found by boosting with these generators; 
$$ 
\psi_I (\phi, \varphi^i, \chi_i, \chi_{i_1i_2}, \ldots )=
e^{\varphi^iJ_{-i}} e^{\phi J_{+-}} e^{ \chi^{i_1i_2}S_{-i_1i_2} }e^{\chi^i S_{+-i}}\ldots \psi_I (0)
\eqno(4.6)$$
\par
Using the commutators of the $E_{11}$ generators with those of the vector representation we find value  of $l_A$ on 
$\psi_I (\phi, \varphi^i, \chi_i, \chi_{i_1i_2}, \ldots )$ and so the momenta $p_A$   it carries. These  momenta are given by 
$$
p_{-}= e^\phi (1 -{1\over 2} \chi^j\chi_j) \sqrt {2}m+\dots  ,\ 
p_i=( -e^\phi \varphi_i (1 -{1\over 2} \chi^j\chi_j) -2\chi_{ij} \chi^j )\sqrt {2}m,\ 
$$
$$
p_{+}= -{1\over 2} \varphi^i\varphi_i (1 -{1\over 2} \chi^j\chi_j)  e^{\phi}   -2 \varphi^k\chi_{kj}\chi^j 
-\chi^{j_1j_2}\chi_{j_1j_2})\sqrt {2}m +\dots+\dots
\eqno(4.7)$$
where $+\dots$ means terms higher than $\chi^2$
and 
$$
p_{-i}= -\chi_ie^\phi \sqrt {2}m+\dots ,\ p_{+-} = \varphi_i\chi^ie^\phi \sqrt {2}m+\dots, \ p_{i_1i_2}=( 2 \varphi_{[i_1} \chi_{i_2]}+2 \chi_{i_1i_2}) \sqrt {2}m+\dots
$$
$$
p_{+i}=(-2 \chi_{ik}\varphi^k - \varphi_i \varphi^j \chi_j e^\phi+{1\over 2} \varphi^j\varphi_j \chi_ie^\phi )\sqrt {2}m+\dots
\eqno(4.8)$$
where $+\dots$ means terms higher than $\chi^1$

\par
The Casimir operator of the Poincare group generalised to  the $I_c(E_{11})$ case is  of the form [59]
$$
L^2\equiv  P_a P^a  + {1\over 2} Z^{a_1a_2}Z_{a_1a_2}+\ldots 
\eqno(4.8)$$
This corrects a typo in equation (4.8) and (4.9)  in reference [32]. We observe that the expression for $L^2$ in equation (4.8) in this paper  is for the generators of the transformations and not for their duals as was given in reference [32].  
\par
There are also an infinite number of other operators which transform into each other under $I_c(E_{11})$ transformations. These are  given by [57]
$$
P_b Z^{ba} , \ Z^{[a_1a_2} Z^{a_3a_4]}+ P_b Z^{b a_1\ldots a_4}, \  P_bZ^{ba_1\ldots a_7}   -3 Z^{[a_1a_2 }Z^{a_3\ldots a_7]} ,\ 
   $$
   $$
    P_c Z^{c a_1 \ldots a_6, b}+ {6.5.3 \over 7}( Z^{b [a_1} Z^{a_2\ldots a_6 ]}-   Z^{ [a_1 a_2} Z^{a_3\ldots a_6] b}) ,\ 
\ldots 
\eqno(4.9)$$
Sine all of these quantities vanish in the rest frame momenta,  and they form an $I_c(E_{11})$ invariant set,  they must vanish on the 
 states  $\psi_I (\phi, \varphi^i, \chi_i, \chi_{i_1i_2}, \ldots )$ of the representation. 
  We note that we could have chosen different rest frame momenta to find another representation and these  may not satisfy the above conditions. In this sense these conditions are a bit like BPS conditions. The reader can check that the momenta of equations  (4.7) and (4.8) do indeed satisfy these conditions. 
\par
The $\phi$ and $\varphi^i$ contain $1+9=10$ degrees of freedom as must be the case as   $p^a$ has 11 degrees of freedom,  but is subject to the condition $p_ap^a=0$. The $\chi^i$ and $\chi^{i_1i_2}$ contain $9+36=45$ degrees of freedom which is as it should be as 
$Z^{a_1a_2}$ has 55 degrees of freedom but it is subject to the 11-1 constraints $p_a p^{ab}=0$
\par
We can transform to x-space by taking the integral 
$$
\tilde \psi( x^a, x_{a_1a_2})= \int d\phi d\varphi^i d\chi^i d\chi^{i_1i_2} \ldots e^{p_a x^a+ p_{a_1a_2 }x^{a_1a_2}+\ldots } 
\psi_I (\phi, \varphi^i, \chi_i, \chi_{i_1i_2}, \ldots )
\eqno(4.10)$$
where the $p_a$, $p^{a_1a_2},\ldots $ are given in equations (4.7) and (4.8). Changing variable to an integral over  momenta we will find the integral takes the generic form  
$$
\tilde \psi( x^a, x_{a_1a_2})= \int dp_a dp^{a_1a_2}  \ldots J \delta (p_ap^a+\ldots =0) \delta (p_a p^{ab}=0)\dots e^{p_a x^a+ p_{a_1a_2 }x^{a_1a_2}+\ldots } 
$$
$$
\psi_I (p_a, p^{a_1a_2} , \ldots )
\eqno(4.11)$$
where $J$ is the Jacobian of the transformation. 
\par
The wavefunction $\tilde \psi$ will obey 
$$
({\partial \over \partial x^a} {\partial \over \partial x^b}\eta^{ab} +{1\over 2}{\partial \over \partial x_{ a_1a_2} }{\partial \over \partial x_{ b_1b_2} }
\eta_{a_1b_1} \eta_{a_2b_2}+\ldots ) \tilde \psi_I (x^A)= 0
\eqno(4.12)$$
as well as 
$$
{\partial \over \partial x^a} {\partial \over \partial x_{ a b} }  \tilde \psi_I (x^A) =0
\eqno(4.13)$$
as well as higher level conditions. 
\par
In the IIA theory at level zero equations (4.12) and (4.13) take the form 
$$
({\partial \over \partial x^a} {\partial \over \partial x^b}\eta^{ab} +{\partial \over \partial y_{ a} }{\partial \over \partial y_b }
\eta_{a b}  ) \tilde \psi_I (x^A)= 0
\eqno(4.14)$$
and 
$$
{\partial \over \partial x^a} {\partial \over \partial y_{ a} }  \tilde \psi_I (x^A) ={1\over 2}\Omega ^{\Pi\Sigma} \partial_{\Pi} \partial_{\Sigma} \tilde \psi_I (x^A)={1\over 2} \partial_{\Pi} \partial^\Pi \tilde \psi_I (x^A)=0
\eqno(4.15)$$
\par
Thus we find that the irreducible representation discussed above,  whose physical states in eleven dimensions are the graviton and the three form, obeys equations  (4.12) and (4.13),  or for the level zero IIA theory,  equations (4.14) and (4.15). While we recognise the first of these as the linearised equation of motion as it appears in the irreducible representation, the second is an example of the weak section condition of equation (1.2). As was noted in reference [32],  the full non-linear E   theory must have some generalisation of both of these equations. 
\par
We will now comment on the dependence of the fields  on the additional coordinates. We see that as  the fields  depend  on $\chi^i$ and $\chi^{i_1i_2},\ldots $ the  additional momenta will be non-zero. As these are the brane charges there must be  branes present.  In particular we can read off  from equation (4.8) the two brane charge $p^{a_1a_2}$. Since a non-trivial dependence on the additional momenta implies a non-trivial dependence on the additional spacetime coordinates we must conclude that the later is due to the presence of branes. This should not be a surprise as the $E_{11}$ symmetry transforms the point particle into the branes and so its   transformations on the irreducible representation will change the dependence on the usual momenta (coordinates) into that on  the additional momenta  (coordinates)  associated with the brane.  
\par
It was argued in reference [37] that the dependence of the fields on the extra coordinates was associated with the presence of branes and in particular the  coordinates could be identified with   the  moduli of the brane solutions arising from large gauge transformations. It would be good to see how the dependence of the fields of the irreducible on the additional coordinates arises from this perspective. 
\par
The first  step to find  the non-linear theory is  to formulate the irreducible representation in a covariant manner by embedding   it into  a linear    representation of $I_c(E_{11})$. This is provided by the fields of the non-linear realisation and the first of the above conditions, equation (4.14) would become the equations of motion of the fields in the linearised theory. The full non-linear equations could be found using the Noether method and, at low levels, these are the equations of motion already known in the non-linear realisation of $E_{11}\otimes_s l_1$. 
\par
It was observed in reference [32] that there would have to be  conditions corresponding to that of equation (4.13) in eleven dimensions  in E theory,  or to equation (4.15) in the IIA theory,   The question we wish to address in this paper is what  should be  these  conditions  in the non-linear theory. It  would seem natural, at least to this author, that they should become, in the non-linear theory,    normal equations,  rather than the  generic section conditions of equations (1.2) and (1.3).


\medskip
{\centerline {\bf 5. IIA at level zero}}
\bigskip
{\bf 5.1 A review}
\medskip
The IIA theory emerges from E theory when we delete node ten in  the $E_{11}$ Dynkin diagram, as shown below. 
$$
\matrix{
& & & & & & & & & & & &  \bullet & 11 & & & \cr 
& & & & & & & & & & & & | & & & & \cr
\bullet & - & \bullet & - & \bullet  &  & \ldots & \ldots  & \bullet & - & \bullet & - & \bullet &-&\bullet&-&\otimes&\cr
1 & & 2 & & 3 & &  &   & 6&  &7  & & 8 & & 9 & & 10 \cr
}
\eqno(5.1.1)$$
At level zero we find the algebra $ SO(10,10) \otimes GL(1)$ whose Caran involution subalgebra is $SO(1,9)\otimes SO(1,9)$, that is, 
$ I_c(SO(10,10) \otimes GL(1))= SO(1,9)\otimes SO(1,9)$.  The non-linear realisation of $E_{11}\otimes_s l_1$, that is, E theory,  was constructed for the IIA theory  level zero in reference [46]. We now   will briefly review this construction and extend it to include the use of Cartan forms,  which is the more usual method used in E theory,  as it will be needed later in this paper. 
\par
 The group element of $E_{11}\otimes_s l_1$  is given, at level zero, by 
$$
g=g_lg_E, \quad {\rm where } \quad g_l= e^{x^a P_a+ y_{\dot a }Q ^{\dot a}} \equiv e^{z^A l_A}
\quad {\rm and } \quad g_E=e^{A_{a_1a_2} R^{a_1a_2}} e^{h_a{}^b K^a{}_b} e^{\tilde \phi \tilde R}
\eqno(5.1.2) $$
The field $h_a{}^b$ and $A_{a_1a_2}$ and $\tilde \phi$ depend on the spacetime coordinates $x^\Pi= (x^a, y_{\dot a})$ while we denote generators of $ SO(10,10) \otimes GL(1)$ by $R^\alpha$. 
\par
The Cartan forms of equation (2.11) can be written as 
$$
{\cal V}_l = dx^\Pi E_\Pi{}^a P_a + dx^\Pi E_\Pi{}_{\dot a} Q^{\dot a} , \quad {\cal V}_A = G_a{}^b \tilde K^a{}_b + H_{a_1a_2} R^{a_1a_2} + G\tilde R
\eqno(5.1.3)$$
\par
Using the commutators of equations (A.3) and (A.4)  in appendix A we find that 
$$
G _{a}{}^b=(e^{-1}d e)_a{}^b,\ \ H_{a_1 a_2}= e_{a_1}{}^{\mu_1}e_{a_2}{}^{\mu_2} dA_{\mu_1 \mu_2}, \ \ G= d\tilde \phi
\eqno(5.1.4)$$
while the vierbein is given by 
$$
{E}_\Pi{}^A= e^{-{\tau\over 2}}\tilde {E}_\Pi{}^A, \quad {\rm where }\quad 
\tilde { E}_\Pi{}^A=  \left(\matrix {e & A(e^{-1})^T\cr
0& (e^{-1})^T\cr}\right)
=\left(\matrix {e_\mu{}^a & A_{\mu\rho} e_b{}^{\rho} \cr
0& e_b {}^\nu \cr}\right)
\eqno(5.1.5)$$
where $e^\tau\equiv  e^{-6\tilde \phi}\equiv e^{-2a} (\det e_\mu {}^a  )$, $a$ is the tachyon field, the matrix $e$ is equal to $e_\mu{}^a$ and $A$ to $A_{\mu\nu}$. In fact $\tilde E_\Pi{}^A$ is  the vierbein, in the vector representation,  corresponding to the SO(10,10) group element $e^{A_{a_1a_2} R^{a_1a_2}} e^{h_a{}^b K^a{}_b}$ and the $e^{-{\tau\over 2}}$ factor rises from the GL(1) part $ e^{\tilde \phi \tilde R}$.  Since  $\det \tilde E_\Pi{}^A=1$ we conclude that   $\det  E_\Pi{}^A= e^{-10\tau}$.
\par
The local tangent space algebra at level zero  is $I_c(SO(10,10))= SO(1,9)\otimes SO(1,9)$ which contains the generators $J_{a_1a_2}$ and $S_{a_1a_2}$.  These are the usual local Lorentz transformation  and  $S^{a_1a_2}= R^{b_1b_2} \eta_{a_1b_1}\eta_{a_2b_2}-R_{a_1a_2}$. Taking $h=1-\tilde \Lambda _{a_1a_2} S^{a_1a_2}$ in equation (2.15 ) we find that  the vierbein transforms as   
 $$
 \delta E_M{}^a= -\tilde \Lambda^{a\dot b}E_M{}_{\dot b}, \quad  \delta E_M{}_{\dot a}= -\tilde \Lambda _{\dot a b} E_M{}^{ b}
 \eqno(5.1.6)$$
  The scalar production that exists in tangent space in the $E_{11}\otimes_s l_1$ non-linear realisation [59]  is given for the tangent 
  space vector  $V^A=(V^a , V_{\dot a} ) $  at level zero by 
 $$
 V^2\equiv V^a V_a + V^{\dot a} V_{\dot a}  = V^A I_{AB} V^B,\quad {\rm where}\quad   V_a=\eta _{ab} V^b, \quad  {\rm and}\quad
 V_{\dot a}= \eta _{\dot a\dot b} V^{\dot b}
 \eqno(5.1.7)$$
where $\eta _{ab}$ and $\eta_ {\dot a\dot b}$ are the Minkowski metrics in ten dimensions. It is clearly   invariant under the transformations of equation (5.1.6) and Lorentz transformations. We raise and lower the tangent  $a,\dot b$ indices with $\eta_{ab}$ and $\eta_{\dot a\dot b}$ and their inverses. 
\par
At level zero $E_{11}$ is just  SO(10,10) and, as discussed in appendix B,  this group has the invariant metric $\Omega_{\theta \Gamma}$ which is quite different to the tangent metric of equation  (5.1.7). This metric will appear in the gauge transformations and can be used to raise and lower world volume indices. It will be important to keep track of the  two different metrics which are  being used in the calculations below. 
 \par
Using equation (A.3) and equation (2.15) we find that the Cartan forms $G_\Pi{}_{,\alpha}$ transform under the local transformation 
$h=1-\tilde \Lambda _{a_1a_2} S^{a_1a_2}$ as 
$$ 
\delta G_a{}^b= \tilde \Lambda^{bc}H_{ca} ,\ \delta H_{a_1a_2}= 2\tilde \Lambda_{[a_1}{}^{c} G_{a_2] c} +d\tilde \Lambda^{a_1a_2}
\eqno(5.1.8)$$
However, as explained, for example in [24,25], a local transformation does not preserve the form of the group element of equation (5.1.2) unless one takes $d\tilde \Lambda^{a_1a_2}-2d\tilde \Lambda^{[a_1 | c}G^c{}_{a_2 ]}$ whereupon 
$$
\delta H_{a_1a_2}= 2\tilde \Lambda_{[a_1}{}^{c} ( G_{a_2] c} + G_{c | a_2] } )= 4 \tilde \Lambda_{[a_1}{}^{c} ( G_{ ( a_2] c )} 
\eqno(5.1.9)$$
\par
The Cartan forms  $G_{\Pi , }{}_{\underline \alpha}$ transforms by  a rigid $E_{11}$ induced 
coordinate transformation  on its $\Pi$ index and by  under a local $I_c(E_{11})$ transformation on its $\underline \alpha$ index. As a result we find that the object $G_{A , } {}_{\underline \alpha} \equiv (E^{-1})_A{}^\Pi G_{\Pi , }{}_{\underline \alpha}$ is inert under the rigid 
$E_{11}\otimes _s l_1$ transformations and only transforms under the $I_c(E_{11})$ transformations as its indices suggest. 
\par
To construct the dynamics of the $E_{11}\otimes_s l_1$ non-linear realisation we generally use  the Cartan forms $G_{\Pi, \alpha}$ of equation (2.11) as they are inert under the rigid $E_{11}$ transformations and  only  transform under the local $I_c(E_{11})$ transformations,  see references [24,25] for the construction of the theory in eleven dimensions. Remarkably the symmetries of the non-linear realisation essentially uniquely determine the dynamics which turns out to be invariant under the usual gauge and general transformations. 
\par
However, for the IIA theory at level zero we took a different approach in reference  [46] and we will 
  use the quantity  $M= g_E I_c(g_E^{-1})$  where $I_c$ is the Cartan involution. Clearly this is inert under  $I_c(E_{11})$ transformations but it transforms under rigid $E_{11}$ transformations as $M\to M^\prime = D(g_0) M D (I_c(g_0^{-1}))$. The symmetries at level zero do not determine the dynamics of the fields at level zero. The reason is that at level zero in the IIA theory  the tangent space algebra, $I_c(E_{11})$  is just the rather humble algebra $SO(1,9)\otimes SO(1,9)$ which is not very  powerful and as a result taking just level zero does not include any of the dual fields and their symmetries. 
  \par
   At level zero in the IIA theory $M$, in the vector representation,    has the form 
$$
M \equiv E E^T, \ {\rm or \ in \ components,}\ M_{\Pi\Sigma}= E_\Pi{}^A I_{AB} E_\Sigma{}^B
\eqno(5.1.10)$$
since $D(g_E)_\Pi{}^A= E_\Pi{}^A$  and $I_c(g^{-1})=g^T$ at level zero as $g\in SO(10,10)$. We have glossed over the precise choice of Cartan involution, we should take the choice that leads to $SO(1,9)\otimes SO(1,9)$. 
\par
Using equation (5.1.4) we find that 
$$
M\equiv{ E}{ E}^T= e^{-\tau} 
\left(\matrix { ee^T- A(e^{-1})^Te^{-1}A&  A(e^{-1})^T e^{-1}\cr 
 (e^{-1})^T\ e^{-1} A   & (e^{-1})^T e^{-1}}\right)
$$
$$
= e^{-\tau} 
\left(\matrix { g_{\mu\nu} -A_{\mu\tau} A^{\tau}{}_\nu & A_\mu{}^\nu \cr 
-A^\mu{}_\nu & g^{\mu\nu}\cr}\right)
\eqno(5.1.11)$$ 
where $g_{\mu\nu}= e_\mu {}^a\eta_{ab} e_\nu{}^b$ and $A_{\mu}{}^{\nu}= A_{\mu\tau} g^{\tau\nu}$ as usual. 
\par
The most general action  which is bilinear in  derivatives and  invariant under the symmetries of the non-linear realisation was found to be given by  [46] 
$$  
A=\int d^{20}z \{c_1 \partial_\Pi    M_{\Gamma \Lambda} \partial_\Sigma   (M^{-1})^{\Gamma \Lambda}(M^{-1})^{\Pi   \Sigma  }
+c_2 \partial_\Pi    (M^{-1})^{\Gamma \Lambda}\partial_\Gamma M_{\Sigma  \Lambda} (M^{-1})^{\Pi   \Sigma  }
$$
$$
+c_3 \partial_\Pi    (M^{-1})^{\Pi   \Xi}\partial_\Gamma (M^{-1})^{\Gamma \Lambda} M_{\Xi \Lambda}
+{c_4\over 400} \partial _\Pi    \det M \partial _\Sigma   (\det M )^{-1}
(M^{-1})^{\Pi   \Sigma  }
$$
$$
+{c_5\over 10}  \det M \partial _\Sigma    (\det M )^{-1} \partial _\Pi   
(M^{-1})^{\Pi   \Sigma  } \}
\eqno(5.1.12)$$ 
where $c_1,\ldots , c_5$ are constants which are not determined by the level zero symmetries. These would be fixed if we were to carry out the calculation at higher levels and in particular to include the  duals of the level zero fields which occur at level two. At level one we find the massless  fields of the Ramond-Ramond sector. 
\par
It will prove useful to express the action of equation (5.8)  in terms of $\tau$ and $\tilde {E}$, which has determinant one, that is, in terms of  $\tau$ and $\tilde M= \tilde E\tilde E^T$. It is given by 
$$  
A=\int d^{20} z  \{e^\tau (e_1 \partial_\Pi    \tilde M_{\Gamma \Lambda} \partial_\Sigma   (\tilde
M^{-1})^{\Gamma \Lambda}(\tilde M^{-1})^{\Pi   \Sigma  }+e_2 \partial_\Pi    (\tilde
M^{-1})^{\Gamma \Lambda}\partial_\Gamma \tilde M_{\Sigma  \Lambda} (\tilde M^{-1})^{\Pi   \Sigma  }
$$
$$
+e_4 e^\tau \partial_\Pi    \tau \partial_\Sigma  \tau 
(\tilde M^{-1})^{\Pi   \Sigma  } 
+e_5e^\tau \partial_\Pi   \tau \partial_\Sigma   (\tilde M^{-1})^{\Pi   \Sigma  }\} 
\eqno(5.1.13)$$
 where $e_1=c_1, e_2=c_2$, $c_3=0$, $e_4=-(20c_1+c_2 +c_4-2c_5)$ and $e_5= -2(c_2- c_5)$. In this last step we set $c_3=0$ as it was found in reference [46 ] that this term was not required. We will take the  coefficients to have the values $e_1= {1\over 4}$, $e_2=-1 $ and $e_4=2=e_5$. 
 \par
 It will be useful to reformulate the theory in terms of the Cartan forms $G_{\Pi}{}_\alpha$, or more precisely in terms of 
 $$
 G_A{}_{,B}{}^C \equiv E_A{}^\Pi G_{\Pi}{}_\alpha (D^\alpha )_B{}^C =  E_A{}^\Pi E_B{}^\Lambda \partial_\Pi E_\Lambda {}^C
 \eqno(5.1.4)$$
We have  not displayed the dots on the relevant indices. 
\par
Using equation (5.1.5) we find that 
$$
G_A{}_{,B}{}^C= E_A{}^\Pi \left (\matrix{ e_b{}^\rho \partial_\Pi e_\rho{}^c & (\partial_\Pi A_{\rho\tau} )e_b{}^\rho e_c{}^\tau \cr
     0& e_\kappa {}^b \partial_\Pi e_c{}^\kappa \cr}\right)
 \eqno(5.1.5)$$
 In terms of the Cartan forms the action of equation (5.1.13)  with the values of the coefficients given below this equation is given by 
 $$
A= \int d^{20} z e^\tau \{  \tilde G_{D,}{}^{ (AB)}( -\tilde G_{D,AB} +4 \tilde G_{A,(BD)}) -4 \nabla^C\tau \tilde G^D{}_{,(CD)}  +2\nabla_C \tau \nabla^C\tau\}
  \eqno(5.1.6)$$
 where $\nabla_C= E_C{}^\Pi \partial_\Pi$ and $\tilde G_{A,B}{}^{C}\equiv \tilde E_A{}^\Pi \tilde E_B{}^\Lambda \partial_\Pi \tilde E_\Lambda {}^C$ is the SO(10,10) Cartan form. Written in the above form the action does not contain the metric $\Omega^{\Pi\Lambda}$ but does contain the above tangent space metric which is used to raise and lower tangent indices. 
 \par
 We close this section by giving some of the properties of the vierbein, more precisely $\tilde E_\Pi {}^A$. As this quantity is a group element  of SO(10,10)  in the vector representation it  obeys the defining condition of this representation, namely 
 $$
 \tilde E_\Pi {}^A \Omega_{AB}= \tilde E^{-1}_B{}^\Gamma  \Omega _{ \Gamma \Pi } 
  \eqno(5.1.7)$$
  In this equation we have explicitly included the superscript ${}^{-1}$ to make it clear that it is the inverse vierbein. One can  then show that the SO(10,10) Cartan form obeys the equation 
  $$
  \tilde G_{\Pi , (A C)} \Omega^{CB}= - \Omega _{AC}  \tilde G_{\Pi , }{}^{(CB)}
    \eqno(5.1.8)$$
where indices on $\tilde G_{\Pi, A}{}^{B} $ are raised and lowed with the tangent space metric. The reader can verify that the explicit  expression for the vierbein of equation (5.1.5) does  indeed obey these equations. 
 \medskip
 {\bf 5.2 Gauge transformations in the IIA theory}
\medskip
The gauge transformation of the vierbein was  given in equation (3.1) and we now evaluate them at level zero in the IIA theory. 
 Using equations (B.6) and (B.7) we find that 
$$
\delta E_\Pi{}^A= {1\over 2} (D^{\Gamma \Xi})_\Pi {}^ \Sigma E_\Sigma{}^ A (D_{\Gamma \Xi})_\Lambda {}^\Delta \partial_\Delta\Lambda^\Lambda +(D^\bullet) _\Pi{}^\Sigma E_\Sigma{}^A (D_\bullet) _\Gamma{}^\Theta \partial _{\Theta }\Lambda^\Gamma 
+ \Lambda ^\Gamma\partial_\Gamma E_\Pi{}^A
$$
$$
=\Omega^{\Xi\Sigma} E_\Sigma{}^A (D_{\Pi \Xi})_\Lambda {}^\Delta \partial_\Delta\Lambda^\Lambda
-{3\over 18} E_\Pi{}^A (D^\bullet) _\Gamma{}^\Theta \partial _{\Theta }\Lambda^\Gamma 
 + \Lambda ^\Gamma\partial_\Gamma E_\Pi{}^A
\eqno(5.2.1)$$
where $(D^\bullet) _\Gamma{}^\Theta$ is the matrix representative for the generator $\tilde R$. Using the equations of appendix A and in particular the equation $C^{-1}{}_{\bullet \bullet }= -{1\over 18}$ we find that 
$$
\delta E_\Pi{}^A= (\partial_\Pi\Lambda^\Lambda - \partial^\Lambda \Lambda_\Pi ) E_{\Lambda}{}^A-{1\over 2}\partial_\Gamma \Lambda^\Gamma E_\Pi{}^A
+ \Lambda ^\Gamma\partial_\Gamma E_\Pi{}^A
\eqno(5.2.2)$$
In this equation we have also  raised and lowered the world indices with the metric $\Omega^{\Pi\Sigma}$ as discussed in appendix B.  In particular $\partial ^\Gamma= \Omega^{\Gamma \Theta} \partial_{\Theta} $ and $ \Lambda _\Gamma= \Omega_{\Gamma \Theta} \Lambda ^\Theta$. 
\par
Taking $E_\Pi{}^A= (E_\mu{}^A , E^{\dot \mu}{}^A)$ and $\Lambda^\Lambda= (\xi^\lambda , \Lambda_{ \lambda} )$, and so 
$\Lambda_{\Pi}= (\Lambda_\mu, \xi^\mu)$ and $\partial^\Pi=(\partial^{\dot \mu} , \partial_\mu)$ , equation (5.2.2),  takes the form 
$$
\delta E_\mu{}^A= \partial_\mu \xi^\lambda E_\lambda{}^A+(\partial_\mu \Lambda_\nu -\partial_\nu \Lambda_\mu) E^{\dot \nu}{}^A
-\partial^{\dot \lambda}\Lambda_\mu E_\lambda{}^A + \xi^\nu\partial_\nu E_\mu{}^A+ \Lambda_{\dot \nu} \partial^{\dot \nu} E_\mu{}^A
-{1\over 2}\partial_\Gamma \Lambda^\Gamma E_\mu{}^A\eqno(5.2.3) $$
$$
\delta E^{\dot \mu}{}^A= \partial^{\dot \mu}\Lambda_\nu E^{\dot \nu}{}^A+(\partial^{\dot \mu}\xi^\lambda- \partial^{\dot \lambda}\xi^\mu) E_\lambda{}^A- \partial_\lambda\xi^\mu E^{\dot \lambda}{}^A+ \xi^\nu\partial_\nu E^{\dot \mu}{}^A+ \Lambda_{\dot \nu} \partial^{\dot \nu} E^{\dot \mu}{}^A
-{1\over 2}\partial_\Gamma \Lambda^\Gamma E^{\dot \mu}{}^A
\eqno(5.2.4) $$
\par
In terms of $\tau$ and $\tilde E_\Pi{}^A$ the local  transformations take the form 
$$
\delta \tilde E_\Pi{}^A= (\partial_\Pi\Lambda^\Lambda - \partial^\Lambda \Lambda_\Pi ) \tilde E_{\Lambda}{}^A
+ \Lambda ^\Gamma\partial_\Gamma \tilde E_\Pi{}^A , \quad 
\delta \tau =\Lambda ^\Gamma\partial_\Gamma \tau + \partial_\Gamma \Lambda^\Gamma 
\eqno(5.2.5)$$
Under the local transformations the Cartan forms transform as 
$$
\delta \tilde G_{C, A}{}^B= \Lambda ^\Gamma\partial_\Gamma  \tilde G_{C, A}{}^B+ \tilde E_C{}^\Pi \partial^\Gamma \Lambda_\Pi \tilde G_{\Gamma , A}{}^B + \tilde E_A^\Gamma \tilde E_\Theta {}^B \tilde E_C{}^\Pi \partial_\Pi ( \partial _\Gamma \Lambda^ \Theta - \partial^\Theta \Lambda _\Gamma) 
\eqno(5.2.6)$$
\par 
In terms of $M$ the gauge variation is given by 
$$
\delta M=  \Lambda ^\Gamma\partial_\Gamma  M+ AM+MA^T- \partial_\Gamma \Lambda^\Gamma M
\eqno(5.2.7)$$
where $A$ is the matrix $A_\Pi{}^ \Lambda\equiv \partial_\Pi\Lambda^\Lambda - \partial^\Lambda \Lambda_\Pi $. 
The gauge  transformations of $\tilde M$ and $\tau$ are given by 
$$
\delta \tilde M=  \Lambda ^\Gamma\partial_\Gamma  \tilde M+ A\tilde M+\tilde MA^T ,\ \delta \tau =  \Lambda ^\Gamma\partial_\Gamma \tau + \partial_\Gamma \Lambda ^\Gamma 
\eqno(5.2.8)$$
Thus we have derived the gauge transformation of the IIA theory at level zero from  the general $E_{11}$ approach, namely equation (3.1). It agrees with that found in Siegel theory [38,39,41] which has been constructed from  $M$. 
\par
It is instructive to work out the local  transformations of the component fields. These are given by the transformations of equation (5.2.3) and  (5.2.4),  but also the transformations of equation (5.1.6). These give the transformation  
$$
\delta \tilde E^{\dot \mu a} =( \partial^{\dot \mu} \xi^{\nu} -  \partial^{\dot \nu} \xi^{\mu} )e_\nu{}^a -\tilde \Lambda ^{a\dot b} e_b{}^{\mu}
\eqno(5.2.9)$$
However, using the local tangent space symmetry to choose the group element to be of Borel form means  that the vierbein of equation (5.1.3) has $\tilde E^{\dot \mu a}=0$.    Using the compensating tangent space transformation to ensure this conditions  requires that 
$\tilde \Lambda ^{a\dot b}=-2 e_\mu{}^a \partial^{[\dot \mu } \xi^{\nu]} e_\nu{}^b$. Using the expression for the component field of the vierbein of equations (5.2.3) and  (5.2.4) and this compensations transformation we find that 
$$
\delta e_\mu{}^a = ( \partial_{ \mu} \xi^{\nu} -  \partial^{\dot \nu} \Lambda_{\mu} )e_\nu{}^a 
+2e_\tau{}^a  \partial^{[\dot \tau } \xi^{\kappa]}A_{\mu\kappa} +\Lambda\cdot \partial  e_\mu{}^a
\eqno(5.2.10)$$
$$
\delta A_{\mu\nu}= 2\partial_{[\mu}\Lambda_{\nu]} +\partial_\mu \xi^\kappa A_{\kappa \nu }+\partial_\nu \xi^\kappa A_{\mu \kappa }
-\partial^{\dot \kappa}\Lambda_\mu A_{\kappa \nu}-\partial^{\dot \kappa}\Lambda_\nu A_{\mu\kappa } 
-2g_{\mu\rho} \partial^{[\dot  \rho}\xi^{\kappa]} g_{\kappa \nu} 
$$
$$
+2 A_{\mu\tau} \partial^{[\dot  \tau }\xi^{\kappa]} A_{\nu \kappa}+\Lambda\cdot \partial A_{\mu\nu}
\eqno(5.2.11)$$
where $\Lambda\cdot \partial = \xi^\mu \partial_\mu + \Lambda_{\dot \mu} \partial^{\dot \mu}$. Thus we recover the known gauge transformations of Siegel theory.


 \medskip
 {\bf 5.3 The spacetime restrictions on the parameters}
 \medskip
In section three we found that the gauge transformations close if equation (3.5) holds. It is instructive to evaluate this condition for the IIA theory at level zero.  The first step is to evaluate $N_{RS}$ as defined in equation (3.6). Using appendix B we find that 
$$
N_{\Delta \Sigma }= (C^{-1})_{\Delta \Sigma  \Theta_1\Theta_2} (D^{\Theta _1\Theta_2})_\Pi{}^\Psi\partial_\Psi\Lambda^\Pi=-{1\over 2} (D_{\Delta \Sigma })_\Pi{}^\Psi\partial_\Psi\Lambda^\Pi
= -{1\over 2}(\partial_\Sigma \Lambda_\Delta- \partial_\Delta\Lambda_\Sigma )
\eqno(5.3.1)$$
where we have raised and lowered indices with the SO(10,10) metric  $\Omega_{\Delta \Sigma }$. The first term in $C_\alpha$ is easiest evaluated by multiplying it by $R^\alpha$ and then,  in IIA theory,  it becomes 
$$
[R^{\Psi_1\Psi_2} , R^{\Xi_1\Xi_2} ] N_{1\Psi_1\Psi_2} N_{2 \Xi_1\Xi_2}= 2 R^{\Pi \Lambda} N_{1 \Pi\Theta} N_{2\Lambda}{}^\Theta   -(1 \leftrightarrow 2) 
\eqno(5.3.2)$$
Taking $\alpha= \Pi \Lambda$ we find that the first term in $C_{\Pi \Lambda}$ is equal to 
$$
{1\over 2} (\partial_\Pi \Lambda _{1\Theta}- \partial_\Theta\Lambda_{1\Pi})(\partial_\Lambda \Lambda _{2}{}^{\Theta}- \partial^\Theta\Lambda_{2\Lambda}) 
 -(1 \leftrightarrow 2) 
 \eqno(5.3.3)$$
We then find the condition of equation (3.5) in the IIA theory, at level zero, is 
$$
C_{\Pi \Lambda}= {1\over 2}( \partial_\Pi\Lambda_{1 \Theta} \partial_\Lambda\Lambda _{2}{}^{\Theta}+ \partial_\Theta\Lambda_{1 \Pi} \partial^\Theta\Lambda _{2 \Lambda} 
 -(1 \leftrightarrow 2) )=0
\eqno(5.3.4)$$
This is not the same as the section conditions of equations (1.2) and (1.3). 
\par
The closure of two transformations of equation (3.2)on  the vierbein is given by 
$$
[\delta_{\Lambda_1} , \delta_{\Lambda_2} ] E_\Pi{}^A=CT+  (D^{\Gamma \Theta})_{\Pi}{}^\Sigma E_\Sigma{}^A C_{\Gamma \Theta}
$$
$$
= CT- ( \partial_\Pi\Lambda_{1 \Theta} \partial^\Lambda\Lambda _{2}{}^{\Theta}+ \partial_\Theta\Lambda_{1 \Pi} \partial^\Theta\Lambda _{2}{}^{ \Lambda} 
 -(1 \leftrightarrow 2) )E_\Lambda{}^A
 \eqno(5.3.5)$$
Where the first term CT is the right-hand side  of equation (5.2.2) with the parameter $\Lambda_c^\Pi=\Lambda_2^\Gamma \partial_\Gamma \Lambda_1^\Pi - \Lambda_1^\Gamma \partial_\Gamma \Lambda_2^\Pi$ and the remaining term contains $C_{\Pi\Lambda}$. 
In terms of components we find that 
$$
[\delta_{\Lambda_1} , \delta_{\Lambda_2} ] E_\mu{}^A=CT 
-(\partial_\mu \xi_2^\tau\partial^{\dot \lambda} \Lambda_{1\tau}+\partial_\mu \Lambda_{2\tau}\partial^{\dot \lambda} \xi_{1}^{\tau}
+\partial^\Gamma \Lambda_{2\mu} \partial_\Gamma \xi^\lambda_1) E_\lambda{}^A
$$
$$
-(\partial_\mu \xi_2^\tau\partial^{\lambda} \Lambda_{1\tau}+\partial_\mu \Lambda_{2\tau}\partial^{ \lambda} \xi_{1}^{\tau}
+\partial^\Gamma \Lambda_{2\mu} \partial_\Gamma \Lambda_{1\lambda}) E^{\dot \lambda}{}^A
\eqno(5.3.6)$$
and 
$$
[\delta_{\Lambda_1} , \delta_{\Lambda_2} ] E^{\dot \mu}{}^A=CT 
-(\partial^{\dot \mu} \xi_2^\tau\partial^{\dot \lambda} \Lambda_{1\tau}+\partial^{\dot \mu} \Lambda_{2\tau}\partial^{\dot \lambda} \xi_{1}^{\tau}
+\partial^\Gamma \xi_{2}{}^{\mu} \partial_\Gamma \xi^\lambda_1) E_\lambda{}^A
$$
$$
-(\partial^{\dot \mu} \xi_2^\tau\partial^{\lambda} \Lambda_{1\tau}+\partial^{\dot \mu} \Lambda_{2\tau}\partial^{ \lambda} \xi_{1}^{\tau}
+\partial^\Gamma \xi_{2}{}^{\mu} \partial_\Gamma \xi_{1}{}^{\lambda}) E^{\dot \lambda}{}^A
\eqno(5.3.7)$$
Since $E^{\dot \mu} {}^a=0$, due to our choice of local subalgebra,  we have to carry out a compensating transformation to find the final gauge transformations. We recall that $\partial_\Gamma \bullet \partial^\Gamma\star= \partial_\mu \bullet \partial^{\dot \mu} \star + \partial^{\dot \mu} \bullet \partial_{ \mu} \star$ for any quantities $\bullet$ and $\star$. 
\par
To better understand the situation we now give a simple account of what are the issues in terms of quantities that are very  familiar. Let us consider just a two form in the presence of gravity. Its transformations under gauge and local translations are 
$$
\delta B_{\mu\nu}= \partial_\mu \Lambda_\nu-  \partial_\nu \Lambda_\mu , \ \delta B_{\mu\nu} = \xi^\Lambda \partial_\lambda B_{\mu\nu} + \partial_\mu \xi^\lambda B_{\lambda \nu} +  \partial_\mu \xi^\lambda B_{\mu\lambda }
\eqno(5.3.8)$$
The closure of two of these transformations can be written in two ways, either 
 $$
[\delta_\Lambda , \delta_\xi ] B_{\mu\nu} = \partial_\mu(\xi^\lambda\partial_\lambda\Lambda_\nu)
-\partial_\nu(\xi^\lambda\partial_\lambda\Lambda_\mu)- \partial_\mu\xi^\lambda \partial_\nu \Lambda_\lambda
+ \partial_\nu\xi^\lambda \partial_\mu \Lambda_\lambda
\eqno(5.3.9)$$
or 
$$
[\delta_\Lambda , \delta_\xi ] B_{\mu\nu}= \partial_\mu(\xi^\lambda\partial_\lambda\Lambda_\nu+\partial_\nu \xi^\lambda \Lambda_\lambda)
-\partial_\nu(\xi^\lambda\partial_\lambda\Lambda_\mu+\partial_\mu \xi^\lambda \Lambda_\lambda)
\eqno(5.3.10)$$
We note that in the absence of gravity $\partial_\mu \xi^\nu=0$ and the two ways of writing the  expressions are the same. We note that the terms of the above two equations occur on the right-hand side of equation (5.3.6) of we take $A=\dot a$  and using  the values of the vierbein of equation (5.1.3). 
\par
Let us first consider the result of equation (5.3.10). This is perfectly good way of proceeding if we just have a two form and gravity. 
However, if the theory has  the extended spacetime $x^\Pi= (x^\mu, y^{\dot \mu})$,  with  the  additional coordinate $y^{\dot \mu}$,   and the diffeomorphisms and gauge transformations appear as a shifts in this spacetime,  $\delta x^\Pi=\Lambda^\Pi= (\xi^\mu, \Lambda_\mu)$ there is a problem. 
\par
We will first  explain how Siegel theory (double field theory) resolves this problem. The closure of two such transformations on the coordinates  leads to a shift in the coordinates by    $\Lambda_c^\Pi=\Lambda_2^\Sigma \partial _\Sigma\Lambda_1^\Pi-(1 \leftrightarrow 2)$. 
The problem is that this is not the closure in equation () where the result of the two transformations has a gauge parameter 
$\xi^\lambda\partial_\lambda\Lambda_\nu+\partial_\nu \xi^\lambda \Lambda_\lambda$ and so should lead to a shift in the coordinate 
$y_{\dot \nu} $ of $\delta y_{\dot \nu}= \xi^\lambda\partial_\lambda\Lambda_\nu+\partial_\nu \xi^\lambda \Lambda_\lambda$. 
While the first term of the  required form,  the second term is not and it implies that all fields $\bullet$ should undergo an additional shift $\delta \bullet = \partial_\nu \xi^\lambda \Lambda_\lambda\partial^{\dot \nu} \bullet$. To get rid of this one can adopt the strong section condition of equation (1.3) which involves a derivative on the field and a derivative on the parameter.  
\par
 Let us carry out the same discussion but  in the general case. The closure of two local transformations is given in equation (5.3.5). The first term contains the a transformations of the vierbein with the parameter $\Lambda_c^\Pi=\Lambda_2^\Gamma \partial_\Gamma \Lambda_1^\Pi - \Lambda_1^\Gamma \partial_\Gamma \Lambda_2^\Pi$ and we now examine the remaining terms one by one beginning with the one that contains the expression $ \partial_\Pi\Lambda_{1 \Theta} \partial^\Lambda\Lambda _{2}{}^{\Theta}$. This terms can be written as 
 $$
= ( \partial_\Pi\Lambda_E^\Lambda- \partial^\Lambda\Lambda_E{}_\Pi) E_\Lambda{}^A
\eqno(5.3.11)$$
Looking at equation (5.2.2) we can identify this term as a partial transformation of the vierbein with an parameter 
$\Lambda_E^\Lambda =-{1\over 2} (\Lambda_{2 \Theta}       \partial^\Lambda\Lambda _{1}{}^{\Theta}-(1\leftrightarrow 2))$. However, this is only the first term in the transformation of the vierbein with this parameter, we also require  in the closure the terms 
$$
\Lambda_E^\Lambda\partial_\Lambda E_\Pi{}^A-{1\over 2} \partial_\Lambda (\Lambda_E^\Lambda)E_\Pi{}^A
\eqno(5.3.12)$$
This vanishes if we use the strong and weak section conditions of equations (1.3) and (1.2) as $ \Lambda_E^\Lambda$ contains the derivative $\partial^\Lambda$. 
\par
The remaining term in the closure of equation (5.3.5) involves the expression $\partial_\Theta\Lambda_{1 \Pi} \partial^\Theta\Lambda _{2}{}^{ \Lambda} $ which also vanishes if we use the section condition of equation (1.3). We note that we  required  generic section conditions as we needed a condition  involving the  parameters and field and another condition which involved  the second one just parameters. 
\par
 We now give the resolution of the same problem from the viewpoint of   this paper. We simply take the closure on the fields to be that on the spacetime, namely with a parameter $\Lambda_2^\Gamma\partial_\Gamma \Lambda_1-\Lambda_1^\Gamma\partial_\Gamma \Lambda_2$ and set the extra terms in the closure to zero namely $C_{\Pi \Lambda}=0$,  or equation (5.3.4). We can do this as the terms involving the usual derivatives are counter balanced by terms that involve derivatives in the additional coordinate $y_{\dot \mu}$.
The condition $C_{\alpha}=0$  is not a generic condition but only restricts the way the spacetime appears in the parameters.  \medskip
 {\bf 5.4 The invariance of the action and further conditions}
 \medskip
 In this section we will discuss the extra constraints that are required to ensure that the theory is invariant under  the  local transformations discussed above.  The result will be  most transparent when working with the action formulation in terms of Cartan forms, namely the one of equation (5.1.6) and the local transformations of equation (5.2.5). The terms in the transformations that involve a shift of coordinates cancel with the term in the variation of $\tau$ of the form $\delta \tau = \partial_\Pi \Lambda^\Pi$. The exception is when the derivatives snag on the parameter of the shift. As a result the variation of the action can be written as 
 $$
\delta A= \int d^{20} z \{ e^\tau G_{D,}{}^{ (AB)}( -2 \delta^\prime G_{D , AB} +8 \delta^\prime G_{A,(BD)} ) 
 -4 \nabla _C e^\tau \delta^\prime G^D{}_{,(C D)} 
$$
$$
+4 e^\tau ( -  G_{D ,}{}^{(C D)} E_C{}^\Pi   + \nabla^C\tau E_C{}^\Pi ) \{ \Omega \partial )^\Gamma (\Lambda\Omega)_\Pi \partial_\Gamma \tau + \partial_\Pi \partial_\Theta\Lambda^\Theta \}\}
  \eqno(5.4.1)$$
  where $\delta^\prime$ is the variation of the Cartan form of equation (5.2.5)  without the shift term and ($\Omega \partial )^\Gamma = \Omega^{\Gamma\Pi} \partial_\Pi = \partial^\Gamma $ and $ (\Lambda\Omega)_\Pi= \Lambda _\Pi$ as before. We write these terms in this way so that the reader can see all the places  where  the metric $\Omega ^{\Theta\Gamma}$ appears. Tangent indices are raised and lowered with the tangent space metric. 
  \par
After a relatively short calculation we find that the variation of the action is given by 
  $$
\delta A=   \int d^{20} z \{ -4 e^\tau G_{\Pi , }{}^{(AB)} E_B{}^\Gamma E_A{} ^\Theta \partial _\Theta ( \Omega \partial )^\Pi (\Lambda\Omega)_\Gamma )
-4 \partial _\Theta \partial_\Pi \tau e^\tau E_C{}^\Theta E_C {}^\Gamma ( \Omega \partial )^\Pi (\Lambda\Omega)_\Gamma
$$
$$
+   e^\tau  G^{D , }{}^{(AB)} (8 E_A{} ^\Gamma G_{\Pi , (BD)} -2 E_D{} ^\Gamma G_{\Pi , (AB)})( \Omega \partial )^\Pi (\Lambda\Omega)_\Gamma ) \}
 \eqno(5.4.2)$$
  The above calculation relies on the fact that the term 
  $$
 \int d^{20} z \{ -4 e^\tau G_{D,}{}^{(A B)} E_D{}^\Gamma E_\Pi{}^B E_A{}^\Theta \partial_\Theta( \Omega \partial )^\Pi (\Lambda\Omega)_\Gamma \}
$$
$$
 =  \int d^{20} z \{ 4 e^\tau E_D{}^\Psi G_{\Psi,}{}^{(B}{}_{ C)}\Omega ^{CA}E_B{}^\Pi E_A{}^ \Theta \partial_\Theta \partial_\Pi (\Lambda\Omega)_\Gamma \}=0
  \eqno(5.4.3)$$
 where we have used equations (5.1.7) and (5.1.8). 
 \par

Since we require an invariant action we need the expression of equation  (5.4.2) to vanish. In Siegel theory it vanishes because this theory adopts from the outset  the section condition of equation (1.3) [38,39,41]. This ensures that every term in equation (5.4.2) vanishes separately. Cleary this is not the only way this expression can vanish. 
\par
The local variation of the action can be written in the form 
 $$
 4 \int d^{20} z \{  e^\tau ( \Omega \partial )^\Pi  G_{\Pi , }{}^{(AB)} E_B{}^\Gamma E_A{} ^\Theta \partial _\Theta (\Lambda\Omega)_\Gamma )
  -   ( \Omega \partial )^\Pi  \partial_\Pi \tau e^\tau E_C{}^\Theta E_C {}^\Gamma\partial _\Theta (\Lambda\Omega)_\Gamma \}+\ldots 
  \eqno(5.4.4)$$
 where $+\ldots$ denotes terms which have more than one derivative acting on the fields. We can rewrite this expression as 
  $$
  \int d^{20} z \{ -2 e^\tau ( \Omega \partial )^\Pi  \partial_{\Pi }( E_A{}^\Gamma E_A{} ^\Theta )
  - 4  ( \Omega \partial )^\Pi  \partial_\Pi  e^\tau E_A{}^\Theta E_A {}^\Gamma \} (\partial_\Theta (\Lambda\Omega)_\Gamma )+\ldots 
  \eqno(5.4.5)$$
 \par
 Given the discussion in section four this is what we would expect namely the variation should vanish if we impose equations of the form 
 $$
 ( \Omega \partial )^\Pi  G_{\Pi , (AB)}+\ldots =0 ,\quad  ( \Omega \partial )^\Pi  \partial_\Pi \tau +\ldots=0
  \eqno(5.4.6)$$
 We stress that these are not generic equations but have a specific  form.
 \par
 We leave it to a future paper to determine what are the precise conditions for the action to be invariant under local transformations, that is, for equation (5.4.2) to vanish. A more systematic  way to proceed would be to calculate the conditions for the equations of motion to be invariant under the local transformations. It would also be good to understand the conditions from E theory from a more abstract viewpoint. 
 \par
We close this section by noting that there is another way the above variation of the action  of equation (5.4.5) could vanish, that is, if 
 $$
  E_A{}^\Theta E_A {}^\Gamma  \partial_\Theta (\Lambda\Omega)_\Gamma +\ldots =0
   \eqno(5.4.7)$$
 If we take terms only linearised in the field, and use  equation (5.15), this leads to the condition 
$$
  h^{\mu\nu} (-\partial_\mu \Lambda_\nu+ \bar \partial_{\dot \mu}\xi_\nu)   
+A^{\mu\nu}  (-\partial_\mu \xi_\nu+ \bar \partial_{\dot \mu}\Lambda_\nu)=0
\eqno(5.4.8)$$
for any $h^{\mu\nu}$ and $A^{\mu\nu}$. It is useful to  recall that $\Lambda^\Gamma=(\xi^\mu, \Lambda_\mu)$ and that $(\Lambda\Omega)_\Gamma=\Lambda_\Gamma=(\Lambda_\mu, \xi^\mu)$.  As a result we conclude that 
$$
\partial_\mu \Lambda_\nu-\bar \partial_\mu\xi_\nu+(\mu \leftrightarrow \nu)=0 ,\ {\rm and } \ 
\partial_\mu \xi_\nu-\bar \partial_{\dot \mu}\Lambda_\nu-(\mu \leftrightarrow \nu)=0 
\eqno(5.4.9)$$
We note that the combination of derivatives and parameters is the opposite to that which occurs in the gauge transformations of the fields which at the linearised level are given by 
$$
\delta h_{\mu\nu}= \partial_\mu \xi_\nu +  \partial_\nu \xi_\mu-(\bar \partial_{\dot \mu} \Lambda_\nu +  \bar \partial_{\dot \nu }\Lambda_\mu) ,\ 
\delta A_{\mu\nu}=  \partial_\mu \Lambda_\nu -   \partial_\nu \Lambda_\mu -( \bar\partial_{\dot \mu} \xi_\nu - \bar\partial_{\dot \nu }\xi_\mu)
\eqno(5.4.10)$$
 As such, these conditions set the parts of the parameters which do not occur in the lowest order  local transformations to zero, so eliminating this redundancy. They do occur in the shift part of the local transformations and so these conditions resolve the discrepancy between the first term and the other two terms in the transformation  of the vierbein of equation (5.2.2). It is this discrepancy that requires the section conditions. In view of this, terms in the action  of the form given in equation (5.4.5) .cannot be cancelled by adding   terms to  the action that is bilinear in the fields. 
\par
Taking a Taylor expansion of the parameters in the extra coordinate we can solve equations(5.4.9) at lowest order  to find 
$$
\xi_\mu (x, y)= \xi_\mu (x) + \hat \xi_{\mu \nu}(x) y^\nu +\ldots , \rm {and }\ 
 \Lambda_\mu  (x, y)= \Lambda_\mu  (x)+\hat  \Lambda_{\mu \nu}(x) y^\nu+\ldots 
 \eqno(5.4.11)$$
provided $\hat \xi_{(\mu \nu)}=\partial_{(\mu} \Lambda_{\nu)}(x)$ and $\hat \Lambda_{[\mu\nu]}=-\partial_{[\mu}\xi_{\nu]}(x)$.
\par
One of the most elementary method to find the restrictions on the dependence on the additional coordinates is to consider how to directly generalise the conditions of equation (4.15), that is $\partial ^\Pi \partial_\Pi\bullet=0 $ where $\bullet$ is any field,  to become   conditions on the covariant fields that are invariant under the local transformations. At the lowest order this requires invariance under the transformations of equation (5.4.10). The reader may verify that the equations 
$$
\bar \partial^{\dot \mu }\partial_{[\mu} A_{\nu_1\nu_2]}+ \partial^{ \mu }\bar \partial_{[\dot \mu} A_{\nu_1\nu_2]}=0
\quad {\rm and }\quad \bar \partial^{[\dot \mu_1}\partial_{[\nu_2} h_{\nu_2]}{}^{\mu_2]}=0
 \eqno(5.4.12)$$
are of the required form and are gauge invariant at lowest order. In a future paper we will hope to find the precise non-linear  equations on the fields, and possibly also on the parameters,  required for the invariance of the action. .


\medskip
{\bf 6. Discussion}
\medskip
In this paper we have studied   the previously proposed [50]  local transformations in E theory. We have shown that they contain the usual diffeomorphism and gauge transformations and  that they form a closed algebra provided we adopt condition (3.5) on the parameters of the transformation.  In section four we studied the dependence of the   irreducible representation of E theory on the coordinates beyond those of the usual spacetime. We showed that they corresponded to the presence of branes in addition to the point particles of the supergravity theories,  in agreement with the considerations [37]. Thus the behaviour of the fields on these additional coordinates is of a non-trivial nature which is in  contrast to that suggested by the section conditions of equation (1.2) and (1.3).
 \par
 In section 5.2 we evaluate the local transformations of E theory, discussed above, in the context of the IIA theory at level zero. 
 This theory has the same fields and spacetime as Siegel theory  and  we find that the local transformations also agree with those found previously in this theory. However, in section 5.3 we evaluate the closure condition of equation (3.5) in this theory and find that it is  not of the form of the section conditions of equations (1.2) and (1.3). Indeed,  it even does not vanish if one uses these conditions. 
\par
It was explained in reference [32], following the same considerations as in section four, that E theory needed conditions on fields and that a straightforward extension of the sections conditions of equations (1.2) and (1.3) were given by the BPS conditions derived in reference [57]. However, given the unusual  nature of the section conditions this author did not pursue  this avenue.  With this in mind   we carry out, in section 5.4,  the variation of the action of Siegel theory under the local transformations and find the condition for it to be invariant. While every term in this expression vanishes if we use the generic section conditions these are clearly not necessary conditions. Indeed, we find that the action will be invariant if we adopt specific equations for the fields of the theory which are those 
suggested by the analysis of the corresponding irreducible representation discussed in section four. However, in this paper, we do not precisely extract these equations from the invariance condition. These new conditions are much more sophisticated than the generic section conditions of equations (1.2) and (1.3) and so in this way one arrives at a new much more interesting theory. 
\par
Section conditions analogous to those of equations (1.2) and (1.3) have also been systematically  used in a modification of E theory  in references [60-62]. The papers [16, 54, 55, 20, 23, 46, 47,  51, 56, 58] constructed parts of E theory from  the low level fields and the  level zero and one coordinates of E theory. The subsequently developed exceptional field theory [63-67]  was constructed from the low level fields and the  level zero and one coordinates of E theory. What was different was that rather than use the symmetries of the non-linear realisation these  authors used local symmetries and section conditions. While some of these have  the generic form of equations (1.2) and (1.3) they also require section conditions  of an even more unusual form involving just one spacetime derivative as well as the  fields.

\medskip
\centerline{\bf Appendix A1. The IIA algebra at level zero}
\medskip
The IIA theory emerges when we take the $E_{11}$ Dynkin diagram and delete node ten as shown below. 
$$
\matrix{
& & & & & & & & & & & &  \bullet & 11 & & & \cr 
& & & & & & & & & & & & | & & & & \cr
\bullet & - & \bullet & - & \bullet  &  & \ldots & \ldots  & \bullet & - & \bullet & - & \bullet &-&\bullet&-&\otimes&\cr
1 & & 2 & &  & &  &   & &  &  & & 8 & & 9 & & 10 \cr
}
\eqno(A.1)$$
At level zero we find the algebra $ SO(10,10) \otimes GL(1)$,  the $ SO(10,10)$ being obvious from the above Dynkin diagram. 
The generators of  $ SO(10,10) \otimes GL(1)$  are given by  
$$
\tilde K^a{}_b,\ R^{ab},\ \ R_{ab} \ \rm{and }\ \tilde R
\eqno(A.2)$$
The generator $\tilde R$ being the $GL(1)$.
\par
The commutators of the $ SO(10,10) \otimes GL(1)$  algebra are  given by 
$$
[\tilde K^a{}_b,\tilde K^c{}_d]=\delta _b^c \tilde K^a{}_d - \delta _d^a
\tilde K^c{}_b,  \ \ 
 [\tilde K^a{}_b, R^{c d}]=\delta_b^c R^{ad}-\delta_b^d R^{ac},
[\tilde K^a{}_b,  R_{c d}]=-\delta_c^a  R_{bd}+\delta_d^a 
R_{bc},
$$
$$
 [R^{ab},  R_{c d}]=\delta_{[c}^{[a}\tilde K^{b]}{}_{d]},\ \ \ [R^{ab}, R^{c d}]=0=[ R_{ab},  R_{c d}]
$$
$$
[\tilde R , R^{ab} ]=0,\ [\tilde R , R_{ab} ]=0,\ [\tilde R , \tilde K^a{}_b ]=0
\eqno(A.3)$$
\par
The level zero part of the vector representation contains the generators $P_a, Q^{\dot a}$ which belong to the 20 representation of 
SO(10,10). They obey the commutators 
$$
[\tilde K^c{}_b, P_a]=-\delta_a^c P_b,\ \ 
 [  R^{ab}, P_c]=- \delta^{[a}_c Q^{\dot b]},\ [ R_{ab},P_c]=0,\ [\tilde R ,P_a]=-3P_a ,
$$
$$
[\tilde K^a{}_b, Q^{\dot c}]=\delta_b^c Q^{\dot a} ,
 \  [  R_{ab},  Q^{\dot c}]= \delta_{[a}^c P_{b]},\ \  [R^{ab},Q^{\dot c}]=0, \   [\tilde R ,Q^{\dot a}]=-3Q^a
\eqno(A.4)$$ 
 \par
The derivation of this IIA algebra from that of $E_{11}$ was given in appendix A of reference [46], 
the identification with the $E_{11}$ generators was given by  
$$
\tilde K^a{}_b = K^a{}_b +{1\over 6} \tilde R, \ \tilde R = 2K^{11}{}_{11} -\sum_{e=1}^{10} K^e{}_e,\  
$$
$$
 R^{a_1a_2}= {1\over 2} R^{a_1a_2 11}, \   R_{a_1a_2}= {1\over 2} R_{a_1a_2 11}, \ Q^{\dot a}= -Z^{a11}
\eqno(A.5)$$
with  $P_a$ the same. 
 Equation (A.16) of that reference contains a misprint  in the $[  R_{ab},  Q^{\dot c}]$ commutator as does equation (A.11) 
 in the  $[R^{ab},  R_{c d}]$  commutator. They are given correctly above. 
\par
Using the invariance of the Killing form we find it to be given by 
$$
C^a{}_{b , }{}^c{}_d\equiv (\tilde K^a{}_b , \tilde K^c{}_d)=\delta^a_d\delta^c_b ,\ 
C^{a_1a_2}{}_{,b_1b_2}\equiv (R^{a_1a_2}, R_{b_1b_2})= 2\delta^{a_1a_2}_{b_1b_2}, \ C\equiv (\tilde R , \tilde R )=-18 
\eqno(A.6)$$
where all  other components vanish. The result for $C$ can not be deduced using the invariance of the Killing form under the level zero  $ SO(10,10) \otimes GL(1)$ algebra but it follows from the more general  $E_{11}$ algebra.  

The inverse Killing form is given by 
$$
(C^{-1})^a{}_{b , }{}^c{}_d =\delta^a_d\delta^c_b ,\ 
(C^{-1})^{a_1a_2}{}_{,b_1b_2}={1\over 2} \delta^{a_1a_2}_{b_1b_2}, \ C^{-1}\equiv =-{1\over 18}
\eqno(A.7)$$
\par
The Cartan involution invariant sub-algebra  of $ SO(10,10) \otimes GL(1)$ has the generators 
$$
J_{ab}= K^c{}_b\eta_{ca}- K^c{}_a\eta_{cb} ,\ S_{ab}= R^{cd}\eta_{ac}\eta_{bd}- R_{ab}
\eqno(A.8)$$
which generate the algebra  $SO(10)\otimes SO(10)$.

\medskip
{\bf Appendix B.  SO(D,D)} 
\medskip
The group  SO(D,D) transforms the $2D$ dimensional vector  $z^A= (x^a, y_{\dot a})$ by $z^\prime = Rz$, or in components 
$z^{A\prime}= R^A{}_B z^B$,  so as to preserves the scalar product $z^T\Omega z= z^A \Omega _{AB} z^B=2x^ay_{\dot a}$ where the metric $\Omega _{AB} $ is given by 
$$
\Omega_{AB}= \pmatrix {0 & I\cr 
                                      I & 0\cr}
\eqno(B.1)$$
where $I$ is the $D\times D$ identity matrix. As a result it obeys the condition 
$$
R^T\Omega+ \Omega R=0 \quad \rm {or\ in \ components } \quad R^C{}_A\Omega_{CB}+ \Omega _{AC} R^C{}_B=0
\eqno(B.2)$$
\par
Lets us write $R^A{}_B$ as 
$$
R^A{}_B=\pmatrix {\tilde K^a{}_b & R^{ab}\cr
                      R_{ab}& \hat K_{\dot a}{}^{\dot b} \cr}
\eqno(B.3)$$
and then equation (B.2) implies that 
$$
R^{ab}=-R^{ba},\ \hat K_{\dot a}{}^{\dot b}=- \tilde K^b{}_a
\eqno(B.4)$$
\par
Carrying out the commutator of two transformations on the vector we find that 
$$
[R^{AB} , R^{CD} ]= \Omega ^{BC} R^{AD}-\Omega ^{AC} R^{BD} -\Omega ^{BD} R^{AC}    +\Omega ^{AD} R^{BC}
\eqno(B.5)$$
Where we lower indices with the metric, for example $R^{A}{}_{B}= R^{AC}\Omega_{CB}$,  and raise with the inverse metric $\Omega^{AB} $  which is, in this case,  numerically equal to $\Omega_{AB}$. Then equation (B.2) becomes $R_{AB}=-R_{BA}$.
\par
We introduce generators $P_A= (P_a, Q^{\dot a})$ in the vector representation which have the commutators 
$$
[ R_{AB} , P_C ]= \Omega_{BC} P_A- \Omega _{AC} P_B\equiv -(D_{AB})_C{}^D P_D
\eqno(B.6)$$
\par
We will need the Killing form of SO(D,D). Using the fact that it is invariant we find it to be 
$$
C^{AB, CD}= -\Omega ^{AC}\Omega ^{BD} +\Omega^{BC}\Omega^{AD} ,\ \rm{and \ its\  inverse}\ 
C_{AB, CD}={1\over 4}( -\Omega _{AC}\Omega _{BD} +\Omega_{BC}\Omega_{AD})
\eqno(B.7)$$
\par
The Cartan involution subalgebra of SO(D,D) is $SO(1,D-1)\otimes SO(1,D-1)$ and it has the generators 
$$ 
J_{{a}}{}_{{b}} \equiv K^{{c}}{}_{{b}}\eta_{{c}{a}} - K^{{c}}{}_{{a}}\eta_{{c} {b}}, \ \ S_{{a}{b}} \equiv 2(R^{{c}{d}}\eta_{{c} {a}} \eta_{{d}{ b}} 
 - {R}_{{a}{b}})\eqno(B.5) $$
The $I_c(SO(D,D))$  algebra is given by 
$$
[J_{{a} {b}}, J_{{c} {d}}]=  \eta_{{b} {c}}J_{{a} {d}}-\eta_{{a} {c}}J_{{b} {d}}-\eta_{{b} {d}}J_{{a} {c}}+\eta_{{b} {c}}J_{{a} {d}} ,\ 
[S_{{a} {b}}, S_{{c} {d}}]=  {1\over 4}(\eta_{{b} {c}}J_{{a} {d}}-\eta_{{a} {c}}J_{{b} {d}}-\eta_{{b} {d}}J_{{a} {c}}+\eta_{{b} {c}}J_{{a} {d}})
$$
$$
[J_{{a} {b}}, S_{{c} {d}}]=  \eta_{{b} {c}}S_{{a} {d}}-\eta_{{a} {c}}S_{{b} {d}}-\eta_{{b} {d}}S_{{a} {c}}+\eta_{{b} {c}}S_{{a} {d}}
\eqno(B.6)$$

\par
Defining  the  generators 
$$
M^{ab}  \equiv {1\over 2} J^{ab}+{1\over 2} S_{ab} ,\ M^{\bar a\bar b}\equiv - {1\over 2} J^{ab}+{1\over 2} S_{ab}
\eqno(B.7)$$
we find the commutators become 
$[M^{ab}, M^{\bar c\bar d}]=0$ , 
while $ M^{ab}$ and $ M^{\bar a\bar b}$ generate the algebra $SO(1,D-1)\otimes SO(1, D-1)$.


\medskip
{\bf Appendix C Alternative variation of the action}
\medskip
 Siegel theory is usually formulated in terms of the matrix $\tilde M_{\Pi\Lambda}$ and so we will now give the variation of the 
  action of equation (5.1.13)  under the local  transformations of equation (5.2.2):
  $$
\int d^{20} x \{4e_1 e^\tau (\partial_\Sigma\partial _\Gamma \Lambda ^\Pi- \partial_\Sigma \partial^\Pi \Lambda_\Gamma ) \tilde M_{\Pi \Lambda } \partial_\Theta 
(\tilde M^{-1})^{\Lambda \Gamma }(\tilde M^{-1})^{\Sigma \Theta }
$$
$$
+2e_1 e^\tau\partial^\Sigma \Lambda_\Pi (\tilde M^{-1})^{\Pi \Theta } \partial_\Sigma  \tilde M _{\Delta\Theta } (\tilde M^{-1} )^{\Delta\Theta }
$$
$$
+ 2e_2 e^\tau \partial^\Pi\Lambda_\Delta (\partial_\Sigma  \tilde M^{-1} \partial_\Pi \tilde M \tilde M^{-1} )^{\Delta\Sigma }
- 2e_2 e^\tau (\tilde M ^{-1} \partial_\Sigma  A \partial _\Pi\tilde M( \tilde M^{-1})^{\Pi \Sigma }
$$
$$
+ 2e_2 e^\tau  \partial_\Sigma  (\tilde M ^{-1} )^{\Pi \Delta} (\partial _\Pi \partial_\Delta\Lambda^\Sigma -\partial_\Pi \partial^\Sigma  \Lambda^\Delta)
$$
$$
 +2e_4 e^\tau (\partial_\Theta \tau \partial_\Sigma \tau \partial^\Sigma  \Lambda_\Pi (\tilde M^{-1})^{\Pi \Theta }+ \partial_\Theta \tau \partial_\Pi\partial_\Delta \Lambda^\Delta  (\tilde M^{-1})^{\Pi \Theta })
$$
$$
+e_5 e^\tau \partial_\Sigma \partial_\Pi \Lambda^\Pi\partial_\Theta  (\tilde M^{-1})^{\Sigma \Theta }
+e_5 e^\tau \partial_\Sigma \tau ( \partial^\Sigma \Lambda_\Delta  (\tilde M^{-1})^{\Delta\Theta } 
+ \partial^\Theta \Lambda_\Delta \partial_\Theta  (\tilde M^{-1})^{\Delta\Sigma } 
$$
$$
-e_5e^\tau \partial_\Sigma \tau \{ (\partial_\Theta  \partial_\Delta \Lambda^\Sigma  -\partial_\Theta \partial^\Sigma \Lambda_\Delta)(\tilde M^{-1})^{\Delta\Theta }
+(\partial_\Theta \partial _\Delta \Lambda^\Theta  - \partial_\Theta \partial^\Theta \Lambda_\Delta)(\tilde M^{-1})^{\Delta\Sigma }\}
\eqno(C.1)$$
 The calculation is simplified by realising that all quantities undergo a shift $\delta \bullet = \Lambda^\Gamma\partial_\Gamma \bullet$ 
where $\bullet$ is any field and so  the action undergoes this shift provided the derivatives do not catch on the parameter $\Lambda$. This shift is   cancelled by a term in  the variation of $e^\tau$. Also  most terms involving $A$ on which no derivative acts cancel.  The variation of the individual terms in the action of equation (5.1.13) can be read off by seeing what are their  coefficients.  
\par
In this calculation  we used the identity 
$$
(\tilde M \partial_\Theta \tilde M^{-1}\Omega)_{\Delta \Gamma }=( \Omega \Omega\tilde M \Omega \partial_\Theta \Omega \tilde M \Omega)_{\Delta \Gamma }
= (\Omega\tilde M^{-1} \partial_\Theta \tilde M)_{\Delta \Gamma }
$$
$$
= (\partial_\Theta  \tilde M \tilde M^{-1} \Omega)_{\Gamma \Delta} = -(\tilde M\partial_\Theta  \tilde M^{-1}\Omega)_{\Gamma \Delta}
\eqno(C.2)$$
where we have used the fact that $\tilde M$ belongs to SO(10,10) and so $\tilde M^{-1}=- \Omega \tilde M\Omega$ and $\Omega^{-1}= \Omega$, see appendix B. 
\par
We will take the coefficients in the action to be given by 
$$
e_1= {1\over 4}, \ e_2=-1,\ e_4=2,\ e_5=2
\eqno(C.3)$$
These values are the ones required for the action to have the usual diffeomorphism and two form gauge symmetries. They are also the ones that lead to the invariance of the action if we were to use the sections constraints of equations (1.2) and (1.3). 
\par
With the coefficients of equation (C.3)  the variation of the action is given by  
$$
2\int d^{20} z \partial^\Theta  \Lambda_\Delta C_\Theta {}^\Delta
\eqno(C.4)$$
where 
$$
C_\Theta {}^\Delta= {1\over 4} e^\tau \partial_\Theta (\tilde M ^{-1})^{\Pi \Lambda } \partial_\Sigma \tilde M _{\Pi \Lambda } ( \tilde M )^{\Delta\Sigma}
-e^\tau (\partial_\Sigma \tilde M^{-1} \partial_\Theta  \tilde M \tilde M^{-1})^{\Delta\Sigma}
$$
$$
- \partial_\Sigma\partial^\Sigma (e^\tau \tilde M^{-1} ) ^{\Delta\Theta } +\partial_\Theta \partial_\Sigma (e^\tau \tilde M ^{-1})^{\Delta\Sigma}
- \partial_\Sigma \partial_\Theta  \tau (e^\tau\tilde M^{-1})^{\Delta\Sigma}
\eqno(C.5)$$

\medskip
{\bf {Acknowledgements}}
\medskip
Peter West wishes to thank Keith Glennon and Josh O'Connor  for discussions, and also the SFTC for support from Consolidated grant Pathways between Fundamental Physics and Phenomenology, ST/T000759/1.
\medskip
{\bf References}
\medskip
{\item{[1]} S.\ Ferrara, J.\ Scherk and B.\ Zumino, 
``Algebraic Properties of Extended Supersymmetry''},
Nucl.\ Phys.\ {\bf B121} (1977) 393;
E.\ Cremmer, J.\ Scherk and S.\ Ferrara, {\it ``SU(4) Invariant
Supergravity Theory''}, Phys.\ Lett.\ {\bf 74B} (1978) 61.
\item{[2]} E. Cremmer and B. Julia,
{\it ``The $N=8$ supergravity theory. I. The Lagrangian''},
Phys.\ Lett.\ {\bf 80B} (1978) 48
\item{[3]} B.\ Julia, {\it ``Group Disintegrations''},
in {\it Superspace \&
Supergravity}, p.\ 331,  eds.\ S.W.\ Hawking  and M.\ Ro\v{c}ek,
Cambridge University Press (1981).
\item{[4]}  B. Julia, in Vertex Operators in Mathematics and
Physics, Publications of the Mathematical Sciences Research
Institute no 3, Springer Verlag 1984. 
\item{[5]} E. Cremmer, B. Julia and J. Scherk, Phys. Lett. 76B
(1978) 409.
\item{[6]} C. Campbell and P. West, {\it ``$N=2$ $D=10$ nonchiral
supergravity and its spontaneous compactification.''} Nucl.\ Phys.\ {\bf B243} (1984) 112.
\item{[{7}]} M. Huq and M. Namazie, {\it ``Kaluza--Klein supergravity in ten dimensions''},
Class.\ Q.\ Grav.\ {\bf 2} (1985).
\item{[{8}]} F. Giani and M. Pernici, {\it ``$N=2$ supergravity in ten dimensions''},
Phys.\ Rev.\ {\bf D30} (1984) 325.
\item{[9]} J, Schwarz and P. West, {\it ``Symmetries and Transformation of Chiral $N=2$ $D=10$ Supergravity''},
Phys. Lett. {\bf 126B} (1983) 301.
\item {[10]} P. Howe and P. West, {\it ``The Complete $N=2$ $D=10$ Supergravity''}, Nucl.\ Phys.\ {\bf B238} (1984) 181.
\item {[11]} J. Schwarz, {\it ``Covariant Field Equations of Chiral $N=2$ $D=10$ Supergravity''}, Nucl.\ Phys.\ {\bf B226} (1983) 269.
\item {[12]} C. Callan, S. Coleman, J. Wess and B. Zumino, {\it Structure of Phenomenological Lagrangians. 1},
Phys.Rev. {\bf 177} (1969) 2239; {\it Structure of phenomenological Lagrangians. 2},  Phys. Rev. {\bf 177} (1969)  2247.
\item{[13]} P. West, {\it $E_{11}$ and M Theory}, Class. Quant. Grav.  {\bf 18}, (2001) 4443, hep-th/ 0104081.
\item{[14]} V. Ogievetsky,  {\it ``Infinite-dimensional algebra of general  covariance group as the closure of the finite dimensional
algebras  of conformal and linear groups"}, Nuovo. Cimento, 8 (1973) 988.
\item{[15]} A. Borisov and V. Ogievetsky, {\it ``Theory of dynamical affine and conformal  symmetries as the theory of the gravitational field"},  Teor. Mat. Fiz. 21 (1974) 329. 
\item{[16]} P. West, {\it $E_{11}$, SL(32) and Central Charges}, Phys. Lett. {\bf B 575} (2003) 333-342,  hep-th/0307098.
\item{[17]} I. Schnakenburg and  P. West, {\it Kac-Moody symmetries of IIB supergravity}, Phys. Lett. {\bf B517} (2001) 421, hep-th/0107181.
\item{[18]} P. West, {\it The IIA, IIB and eleven dimensional theories and their common $E_{11}$ origin}, Nucl. Phys. B693 (2004) 76-102, hep-th/0402140. 
\item{[19]} P. West, {\it Introduction to Strings and Branes}, Cambridge University Press, 2012.
\item{[20]} ÊF. Riccioni and P. West, {\it E(11)-extended spacetime and gauged supergravities}, JHEP {\bf 0802} (2008) 039, ÊarXiv:0712.1795.
\item{[21]} F. Riccioni and P. West, {\it The E(11) origin of all maximal supergravities}, JHEP {\bf 07} (2007)  063,  arXiv:0705.0752.
 \item{[22]} E. Bergshoeff, I. De~Baetselier, and T. Nutma, {\it E(11) and the embedding tensor}, JHEP {\bf 09} (2007)  047,  arXiv:0705.1304.
\item{[23]} A. Kleinschmidt and P. West, {\it Representations of G+++ and the role of space-time},
JHEP 0402 (2004) 033, hep-th/0312247. 
\item{[24]} A. Tumanov and P. West, {\it E11 must be a symmetry of strings and branes },  Phys. Lett. {\bf  B759 }Ê(2016),  663, arXiv:1512.01644. 
\item{[25]} A. Tumanov and P. West, {\it E11 in 11D}, Phys.Lett. B758 (2016) 278, arXiv:1601.03974. 
\item{[26]} K. Glennon and P. West, {\it The non-linear dual gravity equation of motion in eleven dimensions}, Phys.Lett.B 809 (2020) 135714, arXiv:2006.02383.
\item{[27]} A. Tumanov and P. West, {\it E11, Romans theory and higher level duality relations},  Int. J. Mod. Phys. A {\bf 32} (2017) no. 05, 1750023, arXiv:1611.03369.
\item{[28]} L. J. Romans, {\it Massive N=2a Supergravity in Ten-Dimensions},  Phys. Lett. B {\bf 169} (1986)  374.
\item{[29]} M. Pettit and P. West, {\it E theory in seven dimensions}, Int. J. Mod. Phys. A {\bf 34} (2019) no.~25, 1950135,   arXiv:1905.07330.
\item{[30]} P. West,{\it A brief review of E theory}, Proceedings of Abdus Salam's 90th  Birthday meeting, 25-28 January 2016, NTU, Singapore, Editors L. Brink, M. Duff and K. Phua, World Scientific Publishing and IJMPA, {\bf Vol 31}, No 26 (2016) 1630043, arXiv:1609.06863.
\item{[31]} F. Riccioni and P.   West, {\it Dual fields and E(11)}, Phys. Lett. {\bf B 645} (2007)  286,  arXiv:hep-th/0612001.
\item{[32]} P. West,  {\it  Irreducible representations of E theory},  Int.J.Mod.Phys. A34 (2019) no.24, 1950133,  arXiv:1905.07324.
\item{[33]} K. Glennon and P. West, {\it The massless irreducible representation in E theory and how bosons can appear as spinors},  Int. J. Mod. Phys. A {\bf 36} (2021) no.~16, 2150096,   arXiv:2102.02152 [hep-th]
\item{[34]}  P. West,  {\it $E_{11}$ origin of Brane charges and U-duality
multiplets}, JHEP 0408 (2004) 052, hep-th/0406150. 
\item{[35]} P. Cook and P. West, {\it Charge multiplets and masses
for E(11)}, ÊJHEP {\bf 11} (2008) 091, arXiv:0805.4451.
\item{[36]} P. West, {\it Brane dynamics, central charges and $E_{11}$}, JHEP 0503 (2005) 077, hep-th/0412336. 
\item{[37]} P. West,   {\it Spacetime and large local transformations},   Int.J.Mod.Phys.A 38 (2023) 08, 2350045, arXiv:2302.02199. 
\item{[38]} W.  Siegel. {\it Superspace duality in low-energy superstrings}. Phys. Rev. D,48:2826Ð2837, 1993. 
\item{[39]} W.  Siegel. {\it  Two vierbein formalism for string inspired axionic gravity}. Phys.Rev. D, 47:5453Ð5459, 1993. 
\item{[40]} C. Hull and B. Zwiebach, {\it Double Field Theory},  JHEP {\bf 0909} (2009) 099, hep-th/0904.4664.
\item{[41]}  C. Hull and B. Zwiebach, {\it The gauge algebra of double field
theory and Courant brackets}, JHEP {\bf 0909} (2009) 090, hep-th0908.1792.
\item{[42]} O. Hohm, C. Hull and B. Zwiebach, {\it Background independent action for double field theory}, hep-th/1003.5027.
\item{[43]} O. Hohm, C. Hull and B. Zwiebach, {\it Generalised metric
formulation of double field theory},  hep-th/1006.4823. 
\item{[44]} T. Kugo and B. Zwiebach, {\it Target space duality as a symmetry of string field theory},
  Prog.\ Theor.\ Phys.\  {\bf 87}, 801 (1992) hep-th/9201040.
\item{[45]} O. Hohm and S.  Kwak, {\it Frame-like Geometry of Double Field Theory},   J.Phys.A44 (2011) 085404, arXiv:1011.4101. 
\item{[46]}  P. West, {\it E11, generalised space-time and IIA string theory}, Phys.Lett.B 696 (2011) 403-409,   arXiv:1009.2624. 
\item{[47]} A.  Rocen and P. West, {\it E11, generalised space-time and IIA string theory the R-R sector},  in Strings, Gauge fields and the Geometry behind:The Legacy of Maximilian Kreuzer, edited by  Anton Rebhan, Ludmil Katzarkov,  Johanna Knapp, Radoslav Rashkov, Emanuel Scheid, World Scientific, 2013, arXiv:1012.2744. 
\item{[48]} O. Hohm, S. Ki Kwak and B. Zwiebach, {\it Unification of Type II Strings and T-duality}, PhysRevLett.107.171603, arXiv:1106.5452. 
\item{[49]}  I. Jeon, K. Lee and  J. Park, {\it Ramond-Ramond Cohomology and O(D, D) T-duality}, JHEP 09 (2012) 079, arXiv:1206.3478 
\item{[50]} P. West, {\it Generalised Space-time and Gauge Transformations},    JHEP {\bf 08} (2014)  050,    arXiv:1403.6395.
\item{[51]} P. West, {\it Generalised Geometry, eleven dimensions}, JHEP 1202 (2012) 018, arXiv:1111.1642.  
\item{[52]} A. Tumanov and P. West, {\it Generalised vielbeins and non-linear realisations }, JHEP 1410 (2014) 009,  arXiv:1405.7894. 
\item{[53]} I. Schnakenburg and P. West, {\it  Kac-Moody Symmetries of IIB Supergravity}, Phys.Lett. B517 (2001) 421-428, hep-th/0107181. 
\item{[54]}  C. Hillmann, {\it Generalized E(7(7)) coset dynamics and D=11 supergravity}, JHEP 03 (2009) 135, hep-th/0901.1581. 
\item{[55]}   C. Hillmann,  {\it E(7(7)) and d=11 supergravity}   hep-th/0902.1509. 
\item{[56]} P. West, {\it E11, generalised space-time and equations of motion in four dimensions}, {{    JHEP} {\bf 12} (2012)  068},{{   arXiv:1206.7045 [hep-th]}}.
\item{[57]} P. West, {\it Generalised BPS conditions}, Mod.Phys.Lett. A27 (2012) 1250202, arXiv:1208.3397.
\item{[58]} D. Berman, H. Godazgar, M. Perry and P. West, {\it Duality Invariant Actions and Generalised Geometry.}, JHEP 1202 (2012) 108, arXiv:1111.0459.
\item{[59]} M. Pettit and P. West, {\it An E11 invariant gauge fixing}, Int.J.Mod.Phys. A33 (2018) no.01, 1850009, Int.J.Mod.Phys. A33 (2018) no.01, 1850009, arXiv:1710.11024. 
\item{[60]} G. Bossard, A. Kleinschmidt, J. Palmkvist, C. N. Pope, and E. Sezgin, ÒBeyond E11,Ó
JHEP 05 (2017) 020, arXiv:1703.01305 [hep-th]. )
\item{[61]} G. Bossard, A. Kleinschmidt, and E. Sezgin, ÒOn supersymmetric E11 exceptional field
theory,Ó JHEP 10 (2019) 165, arXiv:1907.02080 [hep-th]. 
\item{[62]} G. Bossard, A. Kleinschmidt, and E. Sezgin, ÒA master exceptional field theory,Ó JHEP
06 (2021) 185, arXiv:2103.13411 [hep-th]. 
\item{[63]} O. Hohm and H. Samtleben, Exceptional Form of D=11 Supergravity, Phys. Rev.
Lett. 111 (2013) 231601, arXiv:1308.1673.
\item{[64]} O. Hohm and H. Samtleben, Exceptional Field Theory I: E6(6) covariant Form of
M-Theory and Type IIB, Phys. Rev. D 89, (2014) 066016 , arXiv:1312.0614.
\item{[65]} O. Hohm and H. Samtleben, Exceptional Field Theory II: E7(7) , arXiv:1312.0614.
\item{[66]} H. Godazgar, M. Godazgar, O. Hohm, H. Nicolai, Henning Samtleben, Supersymmet-
ric E7(7) Exceptional Field Theory, arXiv:1406.3235.
\item{[67]} O. Hohm and H. Samtleben, Exceptional Field Theory III: E8(8), Phys. Rev. D 90,
(2014) 066002, arXiv:1406.3348

\end